\documentstyle[12pt]{article}
\begin{document}
\titlepage
\title{ \begin{flushright}
Preprint PNPI-2023,  1994
\end{flushright}
\bigskip\bigskip\bigskip
The calculation of Feynman diagrams in the superstring
perturbation theory}
\date{}
\author{G.S. Danilov\thanks{E-mail address: danilov@lnpi.spb.su}\\
Petersburg Nuclear Physics Institute,\\
Gatchina, 188350, St.-Petersburg, Russia}
\maketitle
\begin{abstract}
The method of the perturbative calculation of the
multi-loop amplitudes in the superstring theories is proposed.
In this method the multi-loop superstring  amplitudes are
calculated from the equations that
are none other than Ward identities. The above equations
are derived from the requirement that the discussed amplitudes
are independent from a choice of gauge of both the {\it  vierbein}
and the gravitino field.  The amplitudes in question are
determined in the unique way by these equations together with the
factorization condition on the multi-loop amplitudes when two handles
move away from each other.  The considered amplitudes are
calculated in the terms of vacuum correlators of superfields
defined on the complex $(1|1)$ supermanifolds. The above
supermanifolds are described by superconformal
versions of Schottky groups. The superconformal Schottky groups
appropriate for this aim are built for all the spinor structures.

Being based only on the gauge invariance together with the
factorization requirement on the multi-loop amplitudes when the
handles move away from each other (the unitarity ), the proposed
method can be used widely in the critical (super)string theories.
Moreover, after an appropriate  modification to be made, this method
can be employed for non-critical (super)string, too. In this
paper  the closed, oriented Ramond-Neveu-Schwarz
superstring is considered, only boson emission amplitudes being
discussed.
The problem of the
calculation of the multi-loop boson emission amplitudes is
concentrated, in mainly, on those spinor structures where
superfields is branched
on the complex $z$-plane where Riemann surfaces are mapped.
In this case the vacuum superfield correlators can not be derived
by a simple extension of the boson string results. The method of the
calculation of the above correlators
is proposed. The multi-loop amplitudes
associated with all the  even spinor structures are  calculated
in the explicit form. A previous discussion of  the divergency
problem is given.

\end{abstract}

\newpage
\section{Introduction}

Superstrings
[1-4]  currently are the only
candidates for a basis of the unified theory including gravity
[5].
Nevertheless, superstrings still require a long way to be
understood enough even in the framework of the perturbation
theory.

It seems that
superstring models do not contain an enormous strong
high energy gravity interaction, which appears in other gravity
theories making them to be non-renormalizable. Moreover, the
one-loop calculations [4,6]
encourages a hope that the unified theory based on superstrings might
be finite.  The problem arises, however, whether the one-loop
approximation results can be extended to all orders of the
perturbation theory. The more so, additional divergences
could appear in multi-loop superstring amplitudes, just as they
do appear [7] in the boson string theory.  These
divergences are due to a degeneration of genus-n Riemann surfaces
$(n>1)$ into a few ones of the lower genus. The essential progress
in the investigation of this problem was achieved [8] recently, but
it seems to be desirable to continue the study of the above problem
in different superstring models.

Furthermore, two essentially different superstring
schemes are presently discussed  [4], they being the
manifestly space-time ( 10-dim.) supersymmetrical Green-Schwarz
scheme [3] and the manifestly world-sheet  supersymmetrical
Ramond-Neveu-Schwarz one [1]. It is generally believed,
however,  that, after the GSO projection [2], the
Ramond-Neveu-Schwarz superstrings also possess a hidden space-time
supersymmetry and, therefore, both the above schemes correspond to
the same physical model.  Till now, however, the general proof
[4] of this statement exists on a quite formal level, the
direct proof being done in the tree and one-loop approximations only
[4].  To prove   the above
statement for multi-loop amplitudes one should verify the validness
of the non-renormalization theorems [4,9].

The study of the discussed problems requires calculations of
the multi-loop amplitudes in question.  Besides, after the
multi-loop amplitudes being calculated, other significant
goals  could be outlined, for example, the perturbative
calculation of ultraviolet and infrared asymptotics of superstring
amplitudes. It might stimulate new ideas beyond the superstring
perturbation theory.

In this paper we present the calculation of the above multi-loop
amplitudes. We employ the
method of the multi-loop calculations in (super)string theories,
which has been proposed in [10-12]. The considered
method allows to obtain the multi-loop amplitudes in the form
appreciable for the investigation of the divergency
problem and of different asymptotics,
as well. Being based only on the gauge invariance together
with the factorization requirement on the multi-loop amplitudes
when the handles move away from each other (the unitarity ),
this method can be used widely in the critical (super)string
theories.  Moreover, after an appropriate  modification to
be made, the above method can be employed for non-critical
(super)string, too.  But in
this paper we concentrate on the closed, oriented
Ramond-Neveu-Schwarz superstring, only boson emission amplitudes
being considered. We calculate in the explicit form  the
multi-loop amplitudes associated with all the   even spinor
structures.  At last, we touch the divergency problem.

The multi-loop calculations in the  superstring theory
are discussed already for a long time.
In the well known scheme [13-16]
the multi-loop Ramond-Neveu-Schwarz amplitudes are
written to be sums over ordinary spin structures [17]
integrated over Riemann moduli.  The above amplitudes are
usually constructed
[13,15,16,18,19] in
the terms of suitable modular forms. In this approach,
however, for more than three loop amplitudes, one is forced to
use complicated sets of moduli that prevents the study of the
amplitudes obtained [20]. Moreover, in this scheme
multi-loop amplitudes appear  to be depended on a choice of
basis of the gravitino zero modes [13,15,16]. It means
that the two-dimensional supersymmetry is lost in the scheme
discussed. Indeed, in the superstring theory both
the "{\it vierbein}" and
the gravitino field are the gauge fields.  Owing to the gauge
invariance the "true" superstring amplitudes are
independent of a choice of a gauge of the above gauge
fields.  Therefore, they have no dependence on a choice of
basis of the gravitino zero modes.

The discussed dependence on a choice of basis of the
gravitino zero modes appears to be a serious difficulty in the
considered
scheme. But one can hope that the above difficulty is absent in
the formulation [21] possessing of the manifest
two-dimensional supersymmetry. It is the formulation, that is
used in the present paper. In this case the multi-loop superstring
amplitudes are obtained [22,23] ( see also
[11,12] ) by the summation over "superspin" structures
integrated over both the even moduli and the odd ones.  The above
superspin structures are defined for superfields  on the
complex $(1|1)$ supermanifolds [21].  They present
supersymmetrical versions of the ordinary spin structures  on
Riemann surfaces. Being twisted about $(A,B)$-cycles,
the superfields are changed by mappings that present
superconformal versions of
fractional linear transformations.  Generally, every considered
mapping depends on $(3|2)$ parameters [21].  For odd
parameters to be arbitrary, the above mappings include, in
addition, fermion-boson mixings. It differs the superspin
structures from the ordinary spin ones. Indeed, the ordinary
spin structures [17] imply that boson fields are
single-valued on Riemann surfaces.  Only fermion fields being
twisted about $(A,B)$-cycles  may receive the factor (-1).
For all odd parameters to be equal to zero  every genus-n
superspin structure $L=(l_1 ,l_2)$ is reduced to the ordinary
$(l_1,l_2)$ spin one.  Here $l_1$ and $l_2$ are the theta function
characteristics:  $(l_1,l_2)=\bigcup_s(l_{1s} ,l_{2s})$ where
$l_{is}\in(0,1/2)$.  The (super)spin structure is even, if
$4l_1l_2=4\sum_{s=1}^ nl_{1s}l_{2s}$ is even.  It is odd, if
$4l_1l_2$ is odd. For  the discussed superspin structures
it is convenient to use superstring analogues of the Schottky
groups [24,25].  Apparently, it is the only modular
parameterization that allows to perform explicit calculations of
the partition functions in the terms of the even and odd moduli.
In the discussed parameterization  fermion
fields are periodical about  the $A_s$-cycle only, if $l_{1s}=0$.
In the  $l_{1s}\neq 0$  case  the fermion fields are
non-periodical about the $A_s$-cycle, superfields being branched
on the complex $z$-plane where Riemann surfaces are mapped.

In the critical superstring theory the problem of the
calculation of the multi-loop boson emission amplitudes is
concentrated, in mainly, on those superspin structures where
at least one of the $l_{1s}$ characteristics is unequal to zero.
Indeed, for superspin structures where all the $l_{1s}$
characteristics are equal to zero,  the  multi-loop amplitudes
can be derived [22] by  a simple extension of the boson
string results [26]. All the other superspin structures
can not be derived in
this way. Generally, the procedure of "sewing" [23]
allows to consider  the discussed superspin structures, but
this scheme seems to be complicated, the results being obtained
in the form that is rather difficult for an investigation.

In the superstring theory the problem of supersymmetrization
of the ordinary spin structures arises.
Generally, there are different ways to supersymmetrize
ordinary spin structures, but do
not all supersymmetrizations appear to be appropriated for the
superstring theory.  Especially, because the space of half-forms
does not necessarily have a basis when there are odd moduli
[27].  Besides, the chosen set of moduli
is due to be appropriate for constructing of supermodular
invariant superstring amplitudes [11,12].
The super-Schottky groups suitable for the superstring
theory have been constructed in [28-30].
For the $l_{1s}$ characteristic to be
equal to zero the super-Schottky groups have been also built
in earlier papers [14,22,23].

In the case when all
the $l_{1s}$ characteristics  are equal to zero, the
superfield vacuum correlators have been derived [11,12,22]
by a simple extension of the boson string
correlators [10,26].  In the
opposite case when at least one of
the $l_{1s}$ characteristics are unequal to zero,
the superfield vacuum correlators cannot be derived
directly from the boson string theory.
The method of calculating the
vacuum superfield correlators assigned to the discussed
superspin structures has been proposed in [28,30].
In details this method is developed in the present paper
where we calculate the vacuum superfield correlators for the
discussed even superspin structures. Together with the results
obtained in [11,12,22] the results of this paper give
the vacuum correlators for all the even superspin structures.
The above correlators are used for the calculation of the
superstring amplitudes by the method developed in this paper.

This method is based on the
path-integral formulation [31] of (super)string theories.
This formulation allows to employ widely  the local gauge
symmetries of the (super)string.  In this case  a considerable
understanding of (super)string theories has been reached [25]
already in the framework of both the (super)conformal gauge
symmetry and the BRST invariant quantization procedure.  But in
the above approach even for the multi-loop boson string
amplitudes it is failed to find [14] factors due to the
moduli volume form and the ghost zero modes, as well.  So  in
the framework of the discussed approach one can hardly hope to
study satisfactory those quite complicated spin structures where
fermion fields appear to be non-periodical about even if the
only $A$-cycle.
The above spin structures can be studied  in the
approach to (super)string theory developed in
[10-12,28]. This  method employs widely not only
the (super)conformal symmetry, but all the local
gauge symmetries of
the (super)string. Besides, the presented method employs
neither BRST quantization nor the bosonization prescription.

In the proposed method [10-12,28]  the multi-loop
amplitudes are calculated from equations that are none other
than Ward identities. The above equations are obtained from
the condition that the discussed amplitudes are independent of
a choice of the gauge fields, for superstrings they being
the {\it vierbein} and the
gravitino field. In particular, multi-loop amplitudes appear to
be independent of a choice of basis of the gravitino zero modes.

The discussed equations are derived in the framework
of the special ghost scheme [10-12] that allows
to calculate  both the moduli volume form and zero mode
contributions by a suitable modification of the vacuum
correlator of the ghost superfields.
Unlike the usual ghost scheme [31], this scheme
includes "global ghost" parameters, as well as ghost fields.
Then, the gauge fields being fixed, the multi-loop amplitudes
are given by integrals over the string and ghost fields together
with these "global ghost" parameters.  As far as the integrals
over the above fields need ultra-violet regularization, the
obtained expressions are used only to derive equations for the
amplitudes in question.

The above equations resemble those discussed in
[14,25].  But, unlike [14,25], these
equations take into account, in addition, the factors due to
both ghost zero modes and the moduli volume form.  It is urgent
especially for those spin structures where fermion fields are
non-periodical about $A$-cycles because in this case the
equations given in [14,25]  have no
solutions at all.
Besides, unlike the equations in [14,25], the
discussed equations are (super)modular invariant.  So, after a
suitable summation over spin structures being performed, one can
be sure that superstring amplitudes  satisfy restrictions due to
the modular invariance though a direct proof of this statement
may be quite difficult.

Being   differential in moduli, the discussed equations
determine the partition functions up to constant
factors.  To calculate all these factors in the terms
of a coupling
constant,  the factorization requirement on the amplitudes
is used when two handles move away from each other.  This
requirement replaces the unitarity equations.
Though the vertices are known  a long
time already, it seems interesting to note that they could be
calculated in this way, too.   So, the
amplitudes turn out to be fully determined by the gauge
invariance together with the "factorization requirement"
above.

For the closed, critical boson string this method gives
[10] the partition functions to be the same, as in
[26].  This approach has been  also applied
[11,12] to calculating the multi-loop boson
emission amplitudes  of closed,
oriented critical Ramond-Neveu-Schwarz superstrings,
superspin structures corresponding to the fermion fields
periodical about all $A$-cycles  being considered.
For the even superspin structures where at least one of the
$l_{1s}$ characteristics is unequal to zero,
the partition functions were previously considered
in  [28]. The above partition
functions were found to be much more complicate than
those obtained
[14,23] by a "naive" extension of the genus-1 ones.
In details the calculation of the discussed
partition functions is considered in the present paper.

This paper is organized as it follows. To explain the method
we consider in Sec.II the closed, critical boson string,
the superstring specification being ignored for the moment.
In Sec.III we give the equations for multi-loop amplitudes
in the superstring theory. In mainly, the above results have
been obtained early [10-12,29], but they are
given in the above Sections because it is necessary for
understanding the following Sections IV-VII.
In Sec.IV we calculate the vacuum
correlator of the scalar superfields for  the even superspin
structures with at least one of the $l_{1s}$ characteristics is
unequal to zero.  Also, in this case we calculate half-forms
and the period matrices assigned to the
supermanifolds [21,32].  In Sec.V we calculate the vacuum
correlator of the ghost superfields for
the above discussed even superspin structures.  In Sec.VI the
formulae for the multi-loop amplitudes associated
with even superspin structures are obtained.  The final
expressions for the discussed
multi-loop amplitudes associated with all the even superpin
structures are given in Section VII. In this Section we also touch
the divergency problem in the closed, oriented superstring theory.
Details of the calculations are given in the Appendices.

\section{Equations for  multi-loop boson string amplitudes}
As it has been noted in Sec.I, we calculate the multi-loop
amplitudes from the equations that are none other than Ward
identities. We start with the multi-loop amplitudes given
in the form of integrals [31] over the
two-dimensional metrics $g^{\alpha\beta}$.
In the boson string theory the above
two-dimensional metrics are the gauge fields.
To write the multi-loop amplitudes in the
form of integrals over these  $g^{\alpha\beta}$ metrics we
map Riemann surfaces on a complex plane $F$ fixing both {\it
kleinian} groups $K_n$ and fundamental domains $\Omega_n$ to be
the same for all genus-n surfaces [10]. Then the
two-dimensional metrics $g^{\alpha\beta}$ could be chosen
in an arbitrary form inside the fundamental domain $\Omega_n$.
All above $g^{\alpha\beta}$ can be reduce to the full
set $\{\hat g^{\alpha\beta}\}$ of the reference metrics by
gauge mappings $\phi$ that are isomorphisms:
$K_n\stackrel{\phi}{\rightarrow} K_n$ and
$\Omega_n\stackrel{\phi}{\rightarrow} \Omega_n$. This reduction
is impossible within the set in question. In the explicit form
\footnote{Throughout this Section the summation
over Greek indexes repeated twice is implied.}
\begin{equation}
g^{\alpha\beta}(F,\overline F)=\partial_\xi
F^\alpha(u,\overline
u)\partial_\eta F^\beta(u,\overline u)\hat g^{\xi\eta}
(u,\overline u,\{\hat q_N\},\{\overline{\hat q_N}\})
\exp{W(u,\overline
u)}\quad.
\end{equation}
In addition to local complex
coordinates $u$ and their complex conjugated $\overline u$, the
references metrics for $n\geq2$  depend also on the set $\{
\hat q_N\}$ of $3n-3$ complex parameters and their complex
conjugated $\overline{\hat q_N}$, the above $\{ \hat q_N\}$
set being defined modulo of the modular group. In other respects
the references metrics are arbitrary.  \footnote{To every set
$\{\hat q_N\}$ one can assign the set of Riemann moduli. As an
example, one can map all the genus-1 surfaces on the rectangle
$(1,ia)$ with $a>0$, $a$ being the same for all the genus-1
surfaces. Then $\hat g^{\alpha\beta}= \hat
g^{\alpha\beta}(u,\overline u,q,\overline q)$ where $q$ is a
complex parameter. If one reduce  $\hat g^{\alpha\beta}$ to
the plane form $(\hat g^{\alpha\beta}\rightarrow
\delta^{\alpha\beta})$ then the surfaces above turn out to be
mapped on quadrangles $(1,\omega)$ with
$\omega=\omega(q,\overline q)$.}   Then every n-loop
amplitude are  rewritten to be the integral over both the
string fields  and the metrics $ g^{\alpha\beta}$ divided by
the (infinite) volume of the gauge
group.  There is no integration over Riemann
moduli because {\it kleinian} groups are the same for all
the genus-n surfaces. Employing eq.(1), one can write
down  $g^{\alpha\beta}$
in the terms of $\hat g^{\alpha\beta}$ (depending on
$\{\hat q_N,\overline{\hat q_N}\}$), as well as the gauge
functions $F^\alpha$ and $W$.  It is worth-while to note that
$g^{\alpha\beta}$ depends not only on functions $F^\alpha$ and
$W$, but on parameters $(\hat q_N,\overline{\hat q_N})$,
too.  So, the integrals over $g^{\alpha\beta}$ turn into the
ones over $(\hat q_N,\overline{\hat q_N})$, as well as
both $F^\alpha$ and $W$.
The region of the integrating over
$(\hat q_N,\overline{\hat q_N})$ is determined by
the modular invariance. Eq.(1) allows to compute the
jacobian of the discussed transformation. After integrating
over $F^\mu$ and $W$,  this jacobian $J$ can be represented by
the integral over ghost fields $c^\mu$ and
$b_{\alpha\beta}=b_{\beta\alpha}$ together with $3n-3$ complex
Grassmann "global ghosts" $\hat\kappa_N$ and their complex
conjugated the $\overline{\hat\kappa_N}$ as
\begin{equation}
J=\int
e^{-S_{gh}}\prod\delta(\sqrt{-\hat g} b_{\alpha\beta}
\hat g^{\alpha\beta})(D b_{\xi\eta}D c^\nu)
\prod_{s=1}^{3n-3}d\hat\kappa_Nd\overline{\hat\kappa_N}\quad.
\end{equation}
In (2) the ghost action $S_{gh}$ is given by
\begin{equation}
S_{gh}=-\int b_{\alpha\beta}\left[P_\mu^{\alpha\beta} c^\mu+
\frac{\partial\hat g^{\alpha\beta}}{\partial \hat q_s}
\hat\kappa_s\right]\sqrt{-\hat g}dud\overline u
\end{equation}
where, in addition to repeated Greek indexes,  the
summation over $s$ is implied. Moreover, $\hat q_s=(\tilde
q_N,\overline{\hat q_N})$ and $\{\hat\kappa_s\}=
\{\hat\kappa_N,\overline{\hat\kappa_N}\}$. Besides,
$P_\mu^{\alpha\beta}$ is the well known [25]
differential operator:
\begin{equation}
P_\mu^{\alpha\beta}=\hat
g^{\alpha\nu}\partial_\nu\delta_\mu^\beta+ \hat
g^{\beta\nu}\partial_\nu\delta_\mu^\alpha-(-\hat g)^{-1/2}
\hat g^{\alpha\beta}\partial_\mu\sqrt{-\hat g}
\hat g^{\alpha\beta}\quad.
\end{equation}
One can see from (4) that
$\hat g_{\alpha\beta} P_\mu^{\alpha\beta}=0$.  It is known
[31] that volume form $(D\hat b_{\xi\eta}D\hat c^\nu)$
in eq.(2) depends on  $\hat g=\det\hat
g_{\alpha\beta}$,  but it will be unessential for
deriving the equations discussed.

The "global ghosts" $\hat\kappa_s$  are the
peculiarity of the presented scheme that differs this
scheme from that developed in [25]. It must be
stressed that in this scheme the
integrating is performed over all modes of the ghost
fields including $b_{\alpha\beta}$-zero modes. The integral
over the above zero modes appears to be finite owing to the
proportional to $\hat\kappa_s$ terms in the ghost action
(3). As a result every n-loop amplitude $A_n^{(b)}$
can be written as the
integral over the string fields and the ghost fields
together with the global Grassmann parameters $\hat\kappa_s$.
But the  integrals over both the ghost
and string fields need ultra-violet regularization that
hampers a direct calculation of $n$-loop amplitudes
$A_n^{(b)}$. So we use the obtained expression for
$A_n^{(b)}$ only to derive Ward
identities
\begin{equation}
\delta_\bot A_n^{(b)}=0
\end{equation}
where
$\delta_\bot A_n$ are alterations of $A_n^{(b)}$ caused by
transverse infinitesimal arbitrary variations $\delta_\bot
g_{\alpha\beta}$ of the references metrics $\hat
g_{\alpha\beta}$ ($\hat g_{\alpha\beta} \delta_\bot
g_{\alpha\beta}=0$ ). Then we employ the above Ward
identities to calculate $A_n^{(b)}$.

For this aim we reduce $\hat g^{\alpha\beta}$ to the
conform plane form by a mapping $u\rightarrow z(u,\bar u)$ on
a new complex plane $z$.
Simultaneously, $\hat\kappa_N\rightarrow
\kappa_N(\{\kappa_M,\overline{\kappa_M}\})$ and
$\hat q_N\rightarrow q_N(\{q_M,\overline{q_M}\})$,
$q_N$ being Riemann moduli. In this case the terms
depending on $\hat\kappa_s$  in  (3) can be
included [10] in
the vector ghost field $r(z,\bar z)$ as
\begin{equation}
r(z,\overline z)=c(z,\overline z)- \sum_N
\left(\frac{\partial z(u,\overline u)} {\partial
q_N}\right)_{u,\overline u}\kappa_N
\end{equation}
where $c$ is the vector conformal   field. The
proportional to $\kappa_N$ terms in (6)
are originated by the proportional
to $\tilde\kappa_N$ terms in eq.(3).
Then the ghost action (3) can be written in
the terms of both the above $r$ field
and the tensor conformal
ghost field $b(z,\bar z)$  as
\begin{equation}
S_{gh}=-2\int(b\overline{\partial}r+\overline
b\partial\overline r)dzd\overline z
\end{equation}
So, $S_{gh}$  has the usual
form [4,25] in the terms of both the tensor conformal
ghost field $b$ and the  vector  field $r$,
 but, unlike [4,25],  the   vector    field in
eq.(7) has  depending on $z$ periods [10]
under {\it kleinian} group transformations
$z\rightarrow g_s(z)$ associated with $2\pi$-twists
about $B_s$-cycles. Indeed, it can
be proved [10] from (6) that
\begin{equation}
r(g_s(z),\overline
g_s(z))=r(z,\overline z)\partial g_s(z)
+\sum_N\left(\frac{\partial g_s(z)}{\partial
q_N}\right)_z\kappa_N \quad.
\end{equation}
Therefore, the ghost vacuum correlator
\begin{equation}
G(z,z')=-<r(z,\overline
z)b(z',\overline{z'})>
\end{equation}
also has periods
on the $z$ plane.  In the explicit form [10]
\begin{equation}
G(g_s(z),z')=(\partial g_s(z))G(z,z')-
\sum_N\left(\frac{\partial
g_s(z)}{\partial q_N}\right)_z \hat\chi_N(z')
\end{equation}
where $\hat\chi_N(z')$ are 2-tensor zero modes:
\begin{equation}
\hat\chi_N(z)=-<\kappa_Nb(z,\overline z)>\quad.
\end{equation}
To avoid misunderstands, it is useful to remind
that in the discussed scheme the integration performs over
all modes of the tensor ghost field including its zero
modes. The integral over the above zero modes appears
to be convergent owing to the proportional
to $\kappa_N$ terms in eq.(6).
Furthermore, eqs.(10) together with the condition
for $G(z,z')$ to be a conform 2-form on $z'$-plane
determine both $G(z,z')$ and
$\hat\chi_N(z')$ in the unique way [10].
Unlike the ghost correlator discussed in [14],
the $G(z,z')$ ghost correlator
satisfying (10) has no unphisical poles [10].

The $n$- loop amplitudes $A_n^{(b)}$ can be written as
\begin{equation}
A_n^{(b)}=
\int Z_n^{(b)}<V>\prod_{N=1}^{3n-3}dq_Nd\overline q_N
\end{equation}
where $Z_n^{(b)}$ is the partition function
and $<V>$  denotes
the vacuum expectation of the vertex product. Then from
eq.(5) one can  obtain [10] the following
equations for $Z_n^{(b)}$:
\begin{equation}
\sum_N\hat\chi_N(z)
\frac{\partial}{\partial q_N}\ln{Z_n^{(b)}}=
-\sum_N\frac{\partial}{\partial q_N}
\hat\chi_N(z)+<T_m^{(b)}+T_{gh}^{(b)}>
\end{equation}
together
with the complex conjugated to (13) ones. In
eqs.(13) the $\hat\chi_N(z')$ tensor
zero modes are the
same as in (10). Furthermore,
$T_{gh}^{(b)}$ and $T_m^{(b)}$ are the stress
tensors of the ghost  and string fields, respectively:
\begin{equation}
T_m^{(b)}=-\frac{1}{2}\partial X^M\partial X_M
\quad{\rm and}\quad
T_{gh}^{(b)}=2b\partial r+(\partial b)r\quad,
\end{equation}
$X^M$ being the string fields; $M=1,2,...,d$ where $d=26$.
One can see that both
$T_{gh}^{(b)}$ and $T_m^{(b)}$ have the usual form [25] in
the terms of the ghost or string fields, but $T_{gh}^{(b)}$ is
calculated with the ghost vacuum correlator, which obey
eqs.(10) instead of that discussed in [14]. So
in the considered scheme the value $T_m^{(b)}+T_{gh}^{(b)}$
is not  conformal 2-form  under {\it
kleinian} group mappings. But the right side of eq.(13),
as well as the left side, appears to be 2-form under
the mapping above. Moreover, it can be prove that eq.(13)
is invariant under  modular transformations
$z\rightarrow\tilde z(z,\{q_N\})$  with the
simultaneous change $q_N\rightarrow \tilde q_N(\{q_M\})$,
$\tilde q_N$ being new moduli.  Furthermore, under the
discussed transformations a
number of rounds about $(A,B)$ cycles corresponds to every
$2\pi$-twist about $B_s$-cycle. Therefore, to every
$z\rightarrow g_s(z)$ mapping one can assign the mapping
$\tilde z\rightarrow \tilde g_{(s)}(\tilde z)$, which describes
the above rounds. So $\tilde z(g_s(z))=\tilde
g_{(s)}(\tilde z)$.  The modular transformations
$G(z,z')\rightarrow\tilde G(\tilde z,\tilde z')$ of the ghost
vacuum correlator  $G(z,z')$ are determined from the
requirement that $\tilde G(\tilde z,\tilde z')$ changes
under {\it kleinian} group
mappings on $\tilde z$-plane in the accordance
with eqs.(10)
written in the terms of new variables. One can verify that
eq.(13) remains invariant under the discussed
transformation. Details of the proof of this statement
are planned to give in an another paper.

Eqs.(13) have been
solved in [10].  The resulting partition functions
appear to be the same as in [26]. It proves the
modular invariance of the discussed partition functions
that has  been not properly proved  in [26].
In the next Section we extend the
above equations (13) to the superstring theory.

\section{Multi-loop Superstring Amplitudes}
In the considered  superstring theory the n-loop
amplitudes $A_n$ are given [22,23] ( see also
[11,12] ) by the sums over "superspin" structures
integrated over $(3n-3|2n-2)$ complex moduli $q_N$ and
their complex conjugated
$\overline q_N$, as well:
\begin{equation}
A_n=\int\prod_N
dq_Nd\overline q_N\sum_{L,L'}\hat Z_{L,L'}^{(n)} <V>_{L,L'}
\end{equation}
where $\hat Z_{L,L'}^{(n)}$ are the partition functions
and $<V>_{L,L'}$ denote the vacuum expectations of the vertex
products. The index $L$ ($L'$)
labels "superspin" structures of right (left) fields.
In this paper we discuss only the even superspin structures.

As it has been already noted in Sec. I, the above
superspin structures are defined for superfields  on the
complex $(1|1)$ supermanifolds [21]. We map these
supermanifolds by the supercoordinate $t=(z|\theta)$ where z is
a local complex coordinate and $\theta$ is its odd partner.
To every discussed supermanifold the period matrix can be
assigned [21,32].
The above genus-$n$ period matrices $\omega(\{q_N\};L)$
present periods about $B$-cycles of   holomorphic
superfunctions forming a suitable basis on  the considered
supermanifold.  It is worth-while to note that the discussed
period matrices depend on the superspin structure in the
terms proportional to odd moduli[23,28,30].

The $q_N$ moduli in (15) are defined modulo the
supermodular group presenting a supersymmetrical version of the
modular one.  For the considered theory to be self-consistent,
the integrand in eq.(15) being multiplied by the product
of the differentials of moduli must be
invariant under transformations of the supermodular group.

Under the discussed transformations
the $t$ supercoordinate is changes by  holomorphic
supersymmetrical mappings [21,32]:
$t\rightarrow\tilde t(t)$. Simultaneously, $q_N\rightarrow \tilde
q_N(\{q_M\})$. Also, generally, the above transformations
turn out the superspin structures into each other:
$L\rightarrow \tilde L$.  Supermodular transformations
$\omega(\{q_N\};L)\rightarrow\omega(\{\tilde q_N\};\tilde L)$
of the period matrices have the same form as the unimodular
transformations of period matrices assigned to Riemann surfaces:
\begin{equation}
\omega(\{q_N\};L)=[A\omega(\{\tilde q_N\};\tilde
L)+B][C\omega(\{\tilde q_N\};\tilde L)+D]^{-1}
\end{equation}
where $A,B,C$ and $D$ are integral matrices discussed in
[9] ( see also [12] ).  In the boson string theory
eqs.(16) would determine in
an implicit form all the new moduli $\{\tilde q\}$ in the terms
of the old ones $\{q\}$ up to arbitrariness due to possible
fractionally linear transformations of Riemann surfaces. To
avoid misunderstands it is necessary to note that in the
superstring theory eqs.(16) are insufficient to
determine all the
$\tilde q_N$ moduli in the terms of the $q_N$ ones because of
the presence of odd moduli.

Two supermanifolds are topological non-equivalent, if they
can not be obtained from each other by a supermodular
transformation. The
region of the integration over even moduli in eq.(15) is
determined by the condition that different varieties of $q_N$ in
(15) correspond to topological non-equivalent
supermanifolds. It is  similar to  the boson string theory where
the region of the integration over moduli is determined by the
modular invariance. Because of the
supermanifods are non-compact in the sense of  ref. [33]
the boundary $\Sigma$ of the discussed region, generally,
depends on the odd moduli. The dependence on odd moduli
in $\Sigma$ must necessarily be taken into account in the
integrating over
the odd moduli in eq.(15).

Every  superspin structure given on a genus-n complex
$(1|1)$  supermanifold is defined by the mappings
$(\Gamma_{a,s}(l_{1s}),\Gamma_{b,s}(l_{2s}))$ that are associated
with rounds about $(A_s,B_s)$-cycles, respectively ($s=1,2,...,n$).
The above mappings present supersymmetrical versions of
fractional  linear transformations.
As it has been noted  in Sec. I, there are different ways to
supersymmetrize the fractional linear transformations,  but do
not all supersymmetrizations appear to be appropriated for the
description of the mapping discussed.  Firstly, the
space of half-forms is due to have a basis. Otherwise [27]
one meets with difficulties in construction of the vacuum
correlator of the scalar superfields.  Besides, the chosen set
of moduli is due to be convenient to distinguish
in eq.(15)
between the superspin srtuctures  $(l_{2s}=0,l_{2s}=1/2)$
and $(l_{1s}=0,l_{2s}=0)$
( see the discussion following after
eq.(19) of this paper ).  The
$(\Gamma_{a,s}(l_{1s}),\Gamma_{b,s}(l_{2s}))$ mappings given
below are conformed with these requirements.

Following [14,22,23] and [11,12] we use for
$\Gamma_{b,s}(l_{2s})$
the superconformal versions of Schottky transformations. The
above Schottky transformations are defined as
\begin{equation}
z\rightarrow\frac{(a_sz+b_s)}{(c_sz+d_s)}\quad{\rm{with}}
\quad a_sd_ s-b_sc_s=1.
\end{equation}
Simultaneously the $\theta$
spinor receives the $(c_sz+d_s)^{-1}$ factor.  Moreover, for
$l_{2s}=0$, the spinors are multiplied by (-1).
The discussed $\Gamma_{b,s}(l_{2s})$ present
superconformal versions of the above transformations.
For $l_{2}=1/2$ one can choose [11,12,13,22,23]
the $\Gamma_{b,s}(l_{2s}=1/2)$ mapping to be
\begin{eqnarray}
z\rightarrow
\frac{a_s(z+\theta\varepsilon_s)+b_s}
{c_s(z+\theta\varepsilon_s)+
d_s},\quad
\theta\rightarrow
\frac{\theta+\varepsilon_s}
{c_s(z+\theta\varepsilon_s)+d_s}\nonumber\\
{\rm{with}}\quad
\varepsilon_s=
\alpha_s(c_sz+d_s)+\beta_s,\qquad a_sd_s-b_sc_s=1-\varepsilon_s
\partial_z\varepsilon_s \quad.
\end{eqnarray}
In (18) the even $( a_s ,b_s ,c_s ,d_s)$
and odd $(\alpha_s,\beta_s)$ parameters can be expressed
[12,13] in the terms of two fixed points
$(u_s|\mu_s)$ and$(v_s|\nu_s)$ on the complex $(1|1)$
supermanifold together with  the multiplier $k_s$ as
\begin{eqnarray}
a=\frac{u-kv-\sqrt k\mu\nu}{\sqrt k(u-v-\mu\nu)},\quad
d=\frac{ku-v-\sqrt k\mu\nu}{\sqrt k(u-v-\mu\nu)},\quad
c=\frac{1-k}{\sqrt k(u-v-\mu\nu)},\nonumber\\
\alpha=(\mu+\sqrt k\nu)(1+\sqrt k)^{-1},\quad
\beta=-(\nu+\sqrt k\mu)(1+\sqrt k)^{-1},
\end{eqnarray}
the index $s$ being omitted.
We choose  $(3|2)$  of the $(u_s,v_s,\mu_s,\nu_s)$
parameters to be the same for all the genus-n
supermanifolds, the rest of them
together with the $k_s$ multipliers being
$(3n-3|2n-2)$ complex moduli $q_N$ in (15). Without
loss of generality one can think that $|k_s|<1$. For the
isomorphism between (18) and (19) to be, we
fix the branch of $\sqrt k_s$,
for example, as $|arg k_S|\leq\pi$.

In fact, the $\arg k_s\rightarrow\arg k_s+2\pi$ replacement
presents the (super)modular transformation turning
$(l_{1s}=0 ,l_{2s}=1/2)$
into $(l_{1s}=0,l_{2s}=0)$.To prove this statement it is
sufficient to check it for the genus $n=1$. For $n=1$
the period $\omega$ is given by [14,25]
$\omega=(2\pi i)^{-1}\ln k$.
So, we see that $\omega$ is turned into $\omega+1$
under the replacement discussed.  Employing the explicit form
of the theta functions, one can verify that this
transformation of $\omega$ is
accompanied  by the replacement $(l_1=0,l_2=1/2)
\rightarrow(l_1=0,l_2=0)$.

We obtain
$\Gamma_{b,s}(l_{2s}=0)$ by the above replacement in (19).
Then the $|\arg k_s|\leq\pi$ condition
provides in eq.(15) the separating of the
$(l_{1s}=0,l_{2s}=1/2)$ and $(l_{1s}=0,l_{2s}=0)$ superspin
structures from each other. It is worth-while to note that under
an another choice of the moduli it might be difficult to separate
between the above superspin structures  in eq.(15) that
should originate difficulties in the calculation of the
superstring amplitudes. As example, one can consider the
transformations given in [22], the $(3|2)$ parameters from
those in eqs.(8) and (9) of ref. [22]  being taken to be
moduli.

Every the transformation (17)
turns the circle $\hat C_s^{(-)}$  into $\hat C_s^{(+)}$ where
\begin{equation}
\hat C_s^{(-)}=\{z:|c_sz+d_s|=1\} \quad{\rm and}\quad
\hat C_s^{(+)}=\{z:|-c_sz+a_s|=1\}.
\end{equation}
The round about the
$\hat C_s^{(-)}$ or $\hat C_s^{(+)}$ circle corresponds to
$2\pi$-twist about $A_s$-cycle.
For $l_{1s}=1/2$ the spinor fields are multiplied by (-1)
under the above round [14]. So,
for $l_{1s}=1/2$ the spinors turn out to be
branched on the complex $z$-plane.
Therefore,
$2\pi$-twists about $A_s$-cycles are associated with the following
$\Gamma_{a,s}^{(o)}(l_{1s})$ mappings:
\begin{equation}
\Gamma_{a,s}^{(o)}(l_{1s})=\{z\rightarrow z,
\theta\rightarrow(-1)^{2l_{1s}}\theta\}.
\end{equation}
The discussed $\Gamma_{a,s}(l_{1s})$ present
superconformal versions of the above
transformations (21).
It is obvious from (21) that
$\Gamma_{a,s}(l_{1s}=0)=I$, but
the $\Gamma_{a,s}(l_{1s}=1/2)$
mapping appears to be non-trivial.

To extend  the discussed mappings (21)
to arbitrary odd moduli it is necessary to find a relation
between odd parameters in $\Gamma_{a,s}(l_{1s}=1/2)$ and
those in $\Gamma_{b,s}(l_{2s})$. For this aim  we employ
[29,30] that for genus $n=1$,
there are no odd moduli.  Indeed, the genus-1 amplitudes
are obtained in the terms of ordinary spin
structures [17]. Then, for
every particular $s$, all the odd parameters in both
$\Gamma_{a,s}(l_{1s})$ and $\Gamma_{b,s}(l_{2s})$ can be reduced
to zero by a suitable transformation $\tilde\Gamma_s$, which is
the same for both the above transformations:
\begin{equation}
\Gamma_{a,s}(l_{1s})=
\tilde\Gamma_s^{-1}\Gamma_{ a ,s}^{(o)}(l_{1s})
\tilde\Gamma_s, \qquad \Gamma_{b,s}(l_{2s})=
\tilde\Gamma_s^{-1}\Gamma_{b ,s}^{(o)}(l_{2s})
\tilde\Gamma_s
\end{equation}
where $\Gamma_{a,s}^{(o)}(l_{1s})$ are given by (21),
$\Gamma_{b,s}^{(o)}(l_{2s})$ are equal to $\Gamma_{b,s}(l_{2s})$
at $\mu_s=\nu_s=0$; $\tilde\Gamma_s$, in addition, depends on
$(\mu_s,\nu_s)$. We choose $\tilde\Gamma_s$ as
\begin{eqnarray}
\tilde\Gamma_s:\qquad
z\rightarrow z_s+\theta_s\tilde \varepsilon_s(z_s),\qquad
\theta\rightarrow\theta_s(1+\tilde\varepsilon_s\tilde
\varepsilon_ s'/2)+
\tilde\varepsilon_s(z_s);\nonumber\\
\tilde\varepsilon_ s'=\partial_{z}\tilde\varepsilon_s(z),
\qquad
\tilde\varepsilon_s(z)=[\mu_s( z-v_s)-
\nu_s( z-u_s](u_s-v_s)^{-1}.
\end{eqnarray}
Employing (23) one can prove that the
transformations (22) remain to be fixed the
supermanifold points $(u_s|\mu_s)$ and $(v_s|\nu_s)$.
In (22)
the $\Gamma_{b,s}(l_{2s})$ mappings appear to be as those
discussed above, $\Gamma_{b,s}(l_{2s}=1/2)$ being given
by (18). Also, one can see from (22) and
(23) that $\Gamma_{a,s}(l_{1s}=1/2)$ is equal to
$\Gamma_{b,s}(l_{2s}=0)$
calculated at $\sqrt k_s=-1$. In the explicit form
\begin{equation}
\Gamma_{ a,s}(l_{1s}=1/2)=
\{z\rightarrow z-2\theta\tilde\varepsilon_s(z),
\quad\theta\rightarrow-\theta(1+2\tilde\varepsilon_s
\tilde\varepsilon_s')+2\tilde\varepsilon_s(z)\}.
\end{equation}
One can see also that, for every $s$, $\Gamma_{ a,s}(l_{1s})$
commutate with $\Gamma_{ b,s}(l_{2s})$. Besides,
\begin{equation}
\Gamma_{b,s}(l_{2s}=0)=
\Gamma_{a,s}(l_{1s}=1/2)\Gamma_{b,s}(l_{2s}=1/2).
\end{equation}
Every $\Gamma_{b,s}l_{2s}$ mapping transforms
the "circle" $C_s^{(-)}$ into the $C_s^{(+)}$ one, the above
"circles" being
\begin{equation}
C_s^{(-)}= \{z :|Q_{\Gamma_{b ,s}}(t)|=1\},\quad
C_s^{(+)}=\{z :|Q_{\Gamma_{b ,s}^{-1}}(t)|=1\}
\end{equation}
where $Q_{\Gamma_{b ,s}}(t)$ is the factor, which the spinor
derivative $D(t)$ receives under the  $\Gamma_{b,s}l_{2s}$
mapping, $D(t)$ being defined as
\begin{equation}
D(t)=\theta\partial_z+\partial_\theta \quad.
\end{equation}
In eq.(27) derivative $\partial_\theta$
is meant to be the "left" one.
For an arbitrary
superconformal mapping $\Gamma=\{t\rightarrow
t_\Gamma=(z_\Gamma(t)|\theta_ \Gamma(t))\}$ this factor
$Q_\Gamma(t)$ is given by
\begin{equation}
Q_\Gamma^{-1}(t)=D(t)\theta_\Gamma(t)\quad;\quad
D(t_\Gamma)= Q_\Gamma(t)D(t).
\end{equation}
Eqs.(26) take into account in the convenient form the
boson-fermion mixing under rounds about $B$-cycles. It is why
we below prefer to use (26) instead of (20)
in the consideration of integrals over $d\theta dz$.
By definition, the exterior ( interior ) of "circles"
(26) is the same as that of  circles
(20).

If $l_{1s}\neq0$,  the cut $\tilde C_s$ appears on the considered
z-plane.  One of its  endcut points  is placed  inside the
$C_s^{(-)}$ circle and the other  endcut point  is placed inside
the $C_s^{(+)}$ one.
Superconformal  p-tensors $F_p(t)$ being considered, every
$\Gamma_{a,s}(l_{1s}=1/2)$ transformation relates $F_p(t)$ with
its value $F_p^{(s)}(t)$ obtained from $F_p(t)$ by $2\pi$-twist
about $\hat C_s^{(-)}$-circle (20).
So, $F_p(t)$ is changed under the $\Gamma_{a,s}(l_{1s})=
\{t\rightarrow
t_s^a\}$ and $\Gamma_{b,s}=\{t\rightarrow t_s^b\}$ mappings as
\begin{equation}
F_p(t_s^a)=F_p^{(s)}(t)Q_{\Gamma_{a,s}}^p(t),\qquad
F_p(t_s^b)=F_p(t)Q_{\Gamma_{b,s}}^p(t).
\end{equation}

The having zero periods vacuum correlator
\begin{equation}
\hat X_{L,L'}(t,\overline t;t',\overline t')=
<X(t,\overline t)X(t',\overline t')>
\end{equation}
of two identical
scalar superfields $X$ can be written
in the terms of the holomorphic Green function
$R_L(t,t')$,   its
periods $J_r(t;L)$ and the period matrix $\omega(\{q_N\};L)=
\{\omega_{sr}(L)\}$ as
\begin{eqnarray}
\hat X_{L,L'}(t,\overline t;t',\overline t')=
\left(J_r(t;L)+
\overline{J_r(t;L')}\right)
[2\pi i(\overline\omega(L')-
\omega(L))^{-1}]_{rs}\times\nonumber\\
\left(J_s(t';L)
+\overline{J_s(t';L')}\right)+
R_L(t,t')+\overline{R_{L'}(t,t')}.
\end{eqnarray}
In (31) an explicit dependence on the moduli
is omitted. Below we also omit the explicit dependence
on $L$ and $L'$ implying that
$\omega\equiv\omega(\{q_N\};L),\omega_{sr}\equiv\omega_{sr}(L),
J_r(t)\equiv J_r(t;L)$ and $R(t,t')\equiv R_L(t,t')$. The above
$R(t,t')$ changes under the  $\Gamma_{a,s}(l_{1s})=\{t\rightarrow
t_s^a\}$ and $\Gamma_{b,s}=\{t\rightarrow t_s^b\}$ mappings as
\begin{equation}
R(t_s^a,t')=R^{(s)}(t,t'),\qquad
R(t_s^b,t')=R(t,t')+J_s(t')\quad.
\end{equation}
Besides,
\begin{equation}
R(t,t')=R(t',t)\pm\pi i
\end{equation}
We normalize $R(t,t')$ by the condition that
\begin{equation}
R(t,t')\rightarrow\ln(z-z'-\theta\theta')\quad
{\rm at}\quad
z\rightarrow z'\quad.
\end{equation}
In this case
\begin{equation}
J_r(t_s^a)=J_r^{(s)}(t)+2\pi i\delta_{rs},
\qquad J_r(t_s^b)=J_r(t)+2\pi i
\omega_{sr}
\end{equation}
where $\omega$ is the period matrix.
The above
$R(t,t')$ is defined up to terms due to scalar zero modes,
which do not contribute into superstring amplitudes. Apart
from an additive constant the discussed zero mode terms can
be fixed by
the condition that Green function $K(t,t')$ being defined as
\begin{equation}
K(t,t')= D(t')R(t,t'),
\end{equation}
decreases to zero when $z\rightarrow\infty$ or
$z'\rightarrow\infty$.
In eq.(36) the spinor derivative
$D(t')$ is defined by  (27).
Then, for even superspin structures,
eqs. (32)-(35) fully determine the
above $K(t,t')$. Then $R(t,t')$ appears to be determined up
to an unessential
additive constant.  It is obvious also that  the discussed
$K(t,t')$ are  $1/2$-supertensors on
$t'$-supermanifold. But on $t$-supermanifold the above
$K(t,t')$ functions have the periods $2\pi i\eta_s(t')$ where
$1/2$-superdifferentials $\eta_s(t')$ are defined as
\begin{equation}
2\pi
i\eta_s(t')=D(t')J_s(t').
\end{equation}

The scheme discussed in Sec.II for the boson string
can be extended
[11,12] to the superstring theory, as well.   In
this case the equations for the partition functions in
(15) are derived from the condition that the
multi-loop amplitudes are independent of a choice of
both the  {\it vierbein}
and the gravitino field. In details the proof of the  above
equations has been given in [11,12],
the final results being given in this paper just below.

In the superstring theory there are 3/2-tensor ghost
superfield $\hat B$ and the vector ghost one $\hat F$.
In the considered scheme [11,12] the above vector
superfield has depending on $t$ periods under rounds about
$(A_s,B_s)$-cycles. For this reason the vacuum correlator
\begin{equation}
G_{gh}(t,t')=<\hat F(t,\overline
t)\hat B(t',\overline t')>
\end{equation}
also has depending on
$t$ periods  under rounds about $(A_s,B_s)$-cycles.
In the explicit form
$$G_{gh}(t_s^a,t')=
Q_{\Gamma_{a,s}}^{-2}(t)\left(G_{gh}^{(s)}(t,t')+
 \sum_N Y_{a,N}^{(s)}(t)\tilde\chi_N(t')\right)\quad,$$
\begin{equation}
G_{gh}(t_s^b,t')=Q_{\Gamma_{b,s}}^{-2}(t)\left(G_{gh}(t,t')+
\sum_N Y_{b,N}^{(s)}(t)\tilde \chi_N(t')\right)
\end{equation}
where
$t_s^a=(g_s^a|\theta_s^a)$ and $t_s^b =(g_s^b|\theta_s^b)$
are the same as in (29) and $\tilde\chi_N$ are 3/2-zero
modes. And $G_{gh}^{(s)}(t,t')$ has been
obtained from $G_{gh}(t,t')$ by $2\pi$-twist
about $\hat C_s^{(-)}$-circle (20).
Furthermore, $Y_{a,N}^{(s)}$ and $Y_{b,N}^{(s)}$ are
polynomials [12] of degree 2 in $(z,\theta)$,  which
are given by
\begin{equation}
Y_{p,N_r}^{(s)}(t)=Y_{p,N_s}(t)\delta_{rs}\quad{\rm
where}\quad p=a,b \quad{\rm and}\quad
Y_{p,N_s}(t)=Q_{\Gamma_{p,s}}^2\left[\frac{\partial g_s^p}
{\partial q_{N_s}}+\theta_s^p\frac{\partial \theta_s^p}
{\partial q_{N_s}}\right].
\end{equation}
As it was explained just after eq.(19), among
$(3n|2n)$ Schottky parameters there are the
$\{q_{N_o}\}$ set of $(3|2)$ parameters, which  are
the same for all the genus-n supermanifolds, and, therefore,
they are not moduli. The sum over $N$ in (39) includes
only those $Y_{p,N_r}^{(s)}(t)$, which associated with the
Schottky parameters that are moduli. But eq.(40) allows
to assign the $Y_{p,N_r}^{(s)}(t)$ polynomial to every the
Schottly parameter including $\{q_{N_o}\}$, too.
As far as $t_s^a=t$ for $l_{1s}=0$, the $Y_{a,N_s}(t)$
polynomials are unequal to zero only, if $l_{1s}\neq0$.
In this case, for every $N_s$, the $Y_{a,N_s}(t)$
function is equal to $Y_{b,N_s}(t)$ calculated
at $\sqrt k_s=-1$. The last statement follows from
eq.(24).  If $\mu_s=\nu_s=0$,
then  among the above $Y_{a,N_s}(t)$ functions
only $Y_{a,\mu_s}(t)$ and $Y_{a,\nu_s}(t)$ are unequal
to zero.\footnote{Throughout this paper
we use for the $\{N_r\}$ indices the same notation
$(k_r,u_r,v_r,\mu_r,\nu_r)$ as for the Schottky parameters.
In this notation, particularly, $q_{k_r}=k_r$,
$q_{u_r}=u_r$ and so one.}
And $Y_{a,k_s}(t)=0$ in any case.  For every $s$, the
$Y_{b,N_s}(t)$ functions form the full set  of the degree-2
polynomials in $(z,\theta)$.  In the terms of Schottky
parameters the $Y_{b,N_s}(t)$ functions have been obtained
in [12],
see also  eqs. (A1) and (A2) in Appendix A of the
present paper.

Eqs.(39) present
the extension of eq.(10) to the superstring theory.
The above equations together with the condition that
$G_{gh}(t,t')$ is 3/2-superconformal form on $t'$-supermanifold
fully determine [11,12] both $G_{gh}(t,t')$
and 3/2-superconformal zero modes $\tilde\chi_N(t')$.
Unlike the ghost correlators discussed in [13,14],
$G_{gh}(t,t')$ satisfying (10), has no unphisical poles
[11,12].

The equations for the
partition functions in (15) have been found to be
[11,12]
\begin{equation}
\sum_N \tilde\chi_N(t)\frac{\partial}{\partial q_N}\ln\hat
Z_{L,L'}^{(n)}=<T_{gh}+T_m>- \sum_N
\frac{\partial}{\partial q_N} \tilde\chi_N(t)
\end{equation}
together with the
equations to be complex conjugated to (41), the 3/2-
zero modes $\tilde\chi_N(t)$ being the same as in (39).
The derivatives with respect to odd moduli in eq.(41)
are implied to be  the "right" ones.
The summation in (41) is performed over all the sets
$\{N_r\}$ of the $(k_r,u_r,v_r,\mu_r,\nu_r)$ indices except
the  $\{N_o\}$ set associated with those Schottky parameters
that are not the moduli.  Furthermore, $T_{gh}$ and $T_m$ are
the stress tensors of the ghost and string superfields,
respectively. In the
explicit form
\begin{equation}
T_m=10(\partial X)D X/2
\end{equation}
where $X$ denotes the scalar superfield, they being
in number 10 ( 10 is the space-time dimension ). Furthermore,
\begin{equation}
T_{gh}=-(\partial F)B-\partial(FB)+D[(DF)B]/2 \quad.
\end{equation}
In  eqs. (42) and (43) the explicit
dependence on the supercoordinate $t=(z|\theta)$ is omitted.
Being written for all superspin structures,  eqs.(41)
form the set of the equation that is invariant under the
supermodular group. The proof of this statement is similar
to that in the boson string theory. We plan to give
this proof in an another paper.

The above $T_m$ and
$T_{gh}$ are calculated in the terms of the vacuum correlators
(31) and (38), the singular at $z\rightarrow z'$
term being removed, as it was explained in [25].  For the
superspin structure $L_0=\bigcup_s(l_{1s}=0 ,l_{2s}=1/2)$  all
the above correlators has been calculated in [11,12].
All they  are obtained by the "naive" supersymmetrization  of
the boson string ones[10]. For the superspin structures
$L(0,l_2)=\bigcup_s(l_{1s}=0,l_{2s})$ all the discussed
correlators are obtained from those for the  $L_0$ superspin
structure by the $\sqrt k_s\rightarrow- \sqrt k_s$ replacement
for every $\sqrt k_s$ associated with $l_{2s}=0$.  If at least
one $l_{1s}\neq 0$ and  odd moduli are
arbitrary, a fermion-boson mixing arises
under twists about both $A$- and $B$-cycles that
prevents to construct superfield vacuum correlators in the
form of Poincar\'e series [14]. The method calculating
the discussed correlators is given below, the even superspin
structures being considered.

\section{Scalar Supermultiplets}
As it has been noted in the  previous Section, the vacuum
correlators (31) of the scalar superfields is
expressed in the terms of the holomorphic Green functions
$R(t,t')$. If all the $l_{1s}$ characteristics are equal to
zero, the above $R(t,t')$ have been obtained
[11,12,22,23,30] by a simple extension
of the boson string result [14,25]. In this Section we
calculate the discussed $R(t,t')$ for those even superspin
structures where at least one of the above $l_{1s}$
characteristics is unequal to zero.

Even for the discussed superspin structures,
there are no special
difficulties in the calculation of $R(t,t')$,
if all the odd parameters are equal to zero.  In this case
$R(t,t')$ is reduced to Green function $R_{(o)}(t,t')$ that
is written in
the terms of both the boson Green function  $R_b(z,z')$ and
the fermion Green one $R_f(z,z')$ as
\begin{equation}
R_{(o)}(t,t')=R_b(z,z')-\theta\theta'R_f(z,z').
\end{equation}
The boson Green function, as well as  its periods $J_{(o)s}$
together with the period matrix $\omega^{(o)}$
have been calculated in [14] ( see also Appendix B of the
present paper ).  The fermion Green function $R_f(z,z')$ in
(44) has been discussed in [18]. We write
$R_f(z,z')$  in the following form:
\begin{equation}
R_f(z,z')=\exp\left\{\frac{1}{2}[R_b(z,z)+R_b(z',z')]-
R_b(z,z')\right\}
\frac{\Theta[l_1,l_2](J|\omega)}{\Theta[l_1,l_2](0|\omega)}
\end{equation}
where Green function $R_b(z,z)$ for $z'=z$ is defined to be
the limit of  $R_b(z,z')-\ln(z-z')$  at $z\rightarrow z'$.
Furthermore, $\Theta$ is the theta function and the symbol $J$
denotes the set of functions
$(J_{(o)s}(z)-J_{(o)s}(z'))/2\pi i$.

Odd moduli being arbitrary, it is not a simple deal to
satisfy eqs.(32) because  in this case fermions are
agitated with bosons under rounds above both $B_s$ and $A_s$
cycles. To solve this problem we built, for  every $s$,
the genus-1 Green function $R_s^{(1)}(t,t')$  that being
twisted  under $(A_s,B_s)$-circles, is changed by
$(\Gamma_{a,s},\Gamma_{b,s})$-mappings (22).
For  odd Schottky parameters ( $\mu_s,\nu_s$ )
to be equal to zero, $R_s^{(1)}(t,t')$ is reduced to
$R_{(o)s}^{(1)}(t,t')$.  Using
eq.(44), one can calculate this
$R_{(o)s}^{(1)}(t,t')$ in the terms of both the genus-1
boson Green function $R_{(b)s}^{(1)}(z,z')$ and the
genus-1 fermion Green function
$R_{(f)s}^{(1)}(z,z';l_{1s},l_{2s})$. The odd parameters
being arbitrary, $R_s^{(1)}(t,t')$ has the following form:
\begin{equation}
R_s^{(1)}(t,t') =R_{(o)s}^{(1)}(t_s,t_s')+
\tilde\varepsilon_s'\theta_s'\Xi_s(\infty,z_s')-
\theta_s\tilde\varepsilon_s'\Xi_s(z_s,\infty)
\quad{\rm for}\quad
s=1,2,...n
\end{equation}
where both  $t_s=(z_s|\theta_s)$,  $t_s'=(z_s'|\theta_s')$ and
$\tilde{\varepsilon}_s'$ are defined by (23).
Furthermore,
$\Xi_s(z,z')$ is
\begin{equation}
\Xi_s(z,z')=(z-z')R_{(f)s}^{(1)}(z,z';l_{1s},l_{2s})
\end{equation}
Apart from two the last terms, $R_s^{(1)}(t,t')$  is obtained by
the $\tilde\Gamma_s$ transformation  of
$R_{(o)s}^{(1)}(t,t')$ that corresponds to eq.(22).
Two the last terms in (46) are proportional to scalar zero
mode on the $t$ or $t'$ supermanifold. They are introduced in
(46) to provide decreasing $K_s^{(1)}(t,t')$ at
$z\rightarrow\infty$ or $z'\rightarrow\infty$
where $K_s^{(1)}(t,t')$ is defined by eq.(36)
for $R= R_s^{(1)}$.

To calculate $R_{(o)s}^{(1)}(t,t')$ for even genus-1
spin structures we use eq.(45)  at $n=1$.
But among even genus-$n$ superspin
structures one  may see  an even number of the handles
associated with the odd genus-1 superspin structure (
$l_{1s}=l_{2s}=1/2$ ).  In this case, because  the
genus-1 fermion zero mode presents,  the genus-1
fermion Green functions $R_{(f)s}^{(1)}(z,z';1/2,1/2)$,
generally, have  periods under twists about both $A_s$
and $B_s$ cycles. And  there is the Green function
among of them, which has the periods only about
$B_s$-cycle. The above Green function is given by
\begin{equation}
R_{(f)s}^{(1)}(z,z')=
\frac{\partial_z\{\Theta[1/2,1/2](J_{(1)}|\omega_s^{(1)})\}}
{\Theta[1/2,1/2](J_{(1)}|\omega_s^{(1)})}
\sqrt{\frac{\partial_{z'}J_{(o)s}^{(1)}(z')}
{\partial_zJ_{(o)s}^{(1)}(z)}}
\end{equation}
where $\Theta$ is
the genus-1 theta
function. Furthermore,
$J_{(1)}=(J_{(o)s}^{(1)}(z)-J_{(o)s}^{(1)}(z'))/2\pi i$ and
$J_{(o)s}^{(1)}$ is the period of $R_{(b)s}^{(1)}(z,z')$,
the period of $J_{(o)s}^{(1)}$ being $2\pi i\omega_s^{(1)}$.
In this case, for every $s$, the  Green function
$K_s^{(1)}(t,t')=D(t')R_s^{(1)}(t,t')$  is changed under
$\Gamma_{b,s}$ transformation as
\begin{eqnarray}
K_s^{(1)}(t,t_s^b(t'))=\left[K_s^{(1)}(t,t')+
\varphi_s(t)f_s(t')\right]Q_{\Gamma_{b ,s}}(t')\nonumber\\
K_s^{(1)}(t_s^b(t),t')=K_s^{(1)}(t,t')+
2\pi i\eta_s^{(1)}(t')-
\varphi_s(t)f_s(t')
\end{eqnarray}
where  $f_s(t')=D(t')\varphi_s(t')$. The above $\varphi_s(t')$
disappears, if $(l_{1s},l_{2s})$-characteristics correspond to
an even genus-1 spin structure. Besides, one can prove that
\footnote{Throughout the paper $\int d\theta\theta=1$}
\begin{equation}
\int\limits_{C_b^{(s)}}f_s(t)\frac{d\theta dz}
{2\pi i}\varphi_s(t) =-1,\quad
\int\limits_{C_b^{(s)}}f_s(t)\frac{d\theta dz} {2\pi i}=0
\quad{\rm and}\quad
\int\limits_{C_b^{(s)}}K_s^{(1)}(t,t')\frac{d\theta' dz'}
{2\pi i}=0.
\end{equation}
In (50) integrating over $z$ is performed along
the contour
$C_b^{(s)}$. On the $z_s$ complex plane (23) this
$C_b^{(s)}$ contour is none other than the $\hat
C_s^{(-)}$ circle (20), see Appendix C for more
details. It is worth- while to note that the last equation of
eqs.(50) remains to be true, if one take $K(t,t')$
instead of $K_s^{(1)}(t,t')$. To prove (50) we start
with the following relations:
\begin{equation}
f_s(t)= -
\int\limits_{C(z)}f_s(t_1)dt_1K_s^{(1)}(t_1,t)\quad{\rm
and}\quad K_s^{(1)}(t,t')=\int\limits_{C(z)}K_s^{(1)}(t,t_1)dt_1
K_s^{(1)}(t_1,t')
\end{equation}
where $dt_1=d\theta_1 dz_1/2\pi i$. In (51) the
infinitesimal contour $C(z)$ gets around $z$-point in the
positive direction. Then we deform this contour to the
$C_s$ contour that surrounds both $C_s^{(-)}$ and $C_s^{(+)}$ circles
(26) together with the $\tilde C_s$ cut  arising for
$l_{1s}\neq0$. In the second of eqs.(51) the
additional term arises  due to the pole at
$z_1=z'$. The above pole contribution is $K_s^{(1)}(t,t')$.
Therefore, in this case the integral about the $C_s$ contour
is equal to zero. Using
eqs.(49), we reduced the  integrals about the $C_s$
contour to those presented in (50).  Besides, from
(51) it is follows ( see Appendix C for more details )
that
\begin{equation}
\int\limits_{C_b^{(s)}}K_s^{(1)}(t,t')\frac{d\theta'
dz'} {2\pi i}\varphi_s(t') =\frac{\varphi_s(t')}{2},\quad
\int\limits_{C_b^{(s)}}f_s(t)\frac{d\theta dz} {2\pi
i}K_s^{(1)}(t,t') =-\frac{f_s(t')}{2}
\end{equation}
The second of
eqs.(52) follows from the first one because
$f_s(t)=D(t)\varphi_s(t)$ and
$D(t)R_s^{(1)}(t,t')=K_s^{(1)}(t',t)$.

It is much more convenient to have deal with
Green function $K(t,t')$ instead of $R(t,t')$
where $K(t,t')$  is defined by (36).
The above Green function
satisfy the following relation:
$$K(t,t')=K_{(o)}(t,t')-
\sum_{r=1}^{n}\int\limits_{C_r}K_{(o)}(t,t_1)
dt_1\delta K_r^{(1)}(t_1,t_2)dt_2K(t_2,t')-
\int\limits_{C_r}
K_{(o)}(t,t_1) \times $$
\begin{equation}
\delta\varphi_r(t_1)dt_1\int
\limits_{C_b^{(r)}}f_r(t_2)dt_2
K(t_2,t')+
\int\limits_{\hat C_r^{(-)}}K_{(o)}(t,t_1)
\varphi_{(o)r}(t_1)dt_1
\int\limits_{C_r}\delta f_r(t_2)dt_2K(t_2,t')
\end{equation}
where $dt=d\theta dz/2\pi i$, etc.  Furthermore,
$K_{(o)}(t,t')$
and $\varphi_{(o)r}$ denote, respectively,
$K(t,t')$ and $\varphi_r$
calculated at all  odd Schottky parameters to be
equal to zero.

In the first integral  on the right side of eq.(53)
the integrations over both $z_1$ and $z_2$ are
performed along $C_r$-contour that
gets around in the positive direction both $C_r^{(-)}$ and
$C_r^{(+)}$ circles (26) together with the
$\tilde C_r$ cut, if this cut presents ( i.e. $l_{1r}\neq0$ ).
Furthermore, $C_b^{(r)}$-contour  is the same as in
(50) and (52), $\hat  C_r^{(-)}$
being defined by (20).

Each of $\delta K_r^{(1)},
\delta\varphi_r$ and $\delta f_r$ in (53) is defined to
be the difference between the corresponding value and that
calculated for zero values of odd Schottky parameters
$(\mu_r,\nu_r)$.  As example, $\delta
K_r^{(1)}(t_1,t_2)=K_r^{(1)}(t_1,t_2)- K_{(o)r}^{(1)}(t_1,t_2)$
where $K_{(o)r}^{(1)}(t_1,t_2)$ is the $K_r^{(1)}(t_1,t_2)$
function taken at $\mu_r=\nu_r=0$.  So the first integral on the
right side could be written down to be the difference of two
integrals, the integrands containing $K_r^{(1)}(t_1,t_2)$ and
$K_{(o)r}^{(1)}(t_1,t_2)$, respectively. In the first of these
integrals the integration over $z_2$ along $C_r$-contour could be
reduced to the integration performed along the $C_b^{(r)}$ one.
The second of these integrals being considered, the integration
over $z_1$ could be reduced to the
integration along $\hat C_r^{(-)}$-contour. After the
above procedure to be made, one can verify that the right
side of (53)  is equal to $K(t,t')$ that proves
eq.(53), for more details
see eqs. (C12) and (C13) in Appendix C.
Only  the odd genus-1 spin structures contribute in two
last terms on the right
side of (53).

The term $K_{(o)}(t,t')=D(t')R_{(o)}(t,t')$ outside the integral
on the right side of (53)
is calculated in the terms of both
$R_b(z,z')$ and $R_f(z,z')$, as it has been explained  above. So,
(53) appears to be an integral equation for $K(t,t')$.
As far as the  kernel of this equation is proportional to odd
parameters, the discussed equation has the unique solution.
The above solution can be obtained by the iteration procedure,
every posterior iteration being, at least, one more power in
odd parameters than a previous one.  Therefore, $K(t,t')$
appears to be a series containing a finite number of
terms. One can verify that the discussed solution possesses
the required properties (32)-(35). Details of
this verification are planned to give in an another paper.

After $K(t,t')$ being determined, the desirable
Green function $R(t,t')$ is calculated without essential
difficulties. It is convenient to determine its periods $J_r$
from the following relation:
\begin{equation}
J_s(t)=\int\limits_{C_s}K(t,t')J_s^{(1)}(t')d\theta'dz'
\end{equation}
where $J_r^{(1)}(t')$ is the period of the genus-1 Green
function $R_r^{(1)}(t,t')$, the period of $J_r^{(1)}(t')$ being
$\omega_r^{(1)}$. The
period matrix $\omega_{rp}$ is given by
\begin{equation}
\omega_{rp}=\omega_r^{(1)}\delta_{rp}+
\int\limits_{C_r}\eta_p(t)J_r^{(1)}(t)d\theta dz=
\omega_r^{(1)}\delta_{rp}+
\int\limits_{C_p}\eta_r(t)J_p^{(1)}(t)d\theta dz
\end{equation}
where $\eta_r(t)$ is defined by (37). In eqs.
(54) and (55) the integration contour
$C_r$ is the same as in (53), the proof of both
(54) and (55) see  in Appendix C. If
all the $l_{1s}$ characteristics are equal to zero, one
can obtain the much more simple formulae
[14,22,23] for the discussed Green functions,
the  $J_s$ functions and the period matrices,
as well.  But it seems that for the rest of the superspin
structures the above values can hardly be obtained in a much more
compact form.  Though the discussed expressions for the above
values seem to be rather complicate, they are quite
appropriate for investigating almost degenerated surfaces.
Therefore, one can hope, at least, to study the divergency
problem for the superstring amplitudes.

In addition to (53), we  obtain  some
another equations that will be convenient in Sec.VI to
calculate  the partition functions. For this aim we start
with the representations of $K(t,t')$ in the form of the
integrals along the infinitesimal contour
$C(z')$ that is the same as in (51), the integrands
being $K_s^{(1)}(t,t_1)K(t_1,t')$ where $s=1,2,...n$. The above
$C(z')$ contour being deformed, the discussed relations   are
transformed into the following ones:
\begin{equation}
K(t,t')=K_s^{(1)}(t,t')+\sum_{r\neq s}
\int\limits_{C_r}K_s^{(1)}(t,t_1)
\frac{d\theta_1dz_1} {2\pi i}K(t_1,t')-\varphi_s(t)\int
\limits_{C_b^{(s)}}f_s(t_1)\frac{d\theta_1dz_1}
{2\pi i}K(t_1,t')
\end{equation}
where  $C_r$-contours are the same as in (53)  and
$s=1,2,...n$. There is no the term with $r=s$ in the sum over
$r$ because the discussed term is given by the integral along
$C_b^{(s)}$- contour on the right side of (56),
$C_b^{(s)}$- contour being the same as in (50).
Indeed, as far as the $\Gamma_{a,s}$ and $\Gamma_{b,s}$ mappings
are the same for both $K_s^{(1)}$ and $K$, the discussed
integral along the $C_s$ contour  can be reduced to the one
along $C_b^{(s)}$. The difference of the  resulting integral
and the last term in eq.(56)
appears to be
\begin{equation}
-\int\limits_{C_b^{(s)}}\left[K_s^{(1)}(t,t'')+\varphi_s(t)f_s(t'')\right]
\frac{d\theta'' dz''}{2\pi i}2\pi i\eta_s(t')
\end{equation}
that is equal to zero because of (50) and  (52).

Every  $C_r$-contour being considered, we denote
K(t,t')  for $z\in C_r$  as $\tilde K_r(t,t')$. We use
(56)  to calculate
the set $\{\tilde K\}$ of the above $\tilde K_r$  functions.
As soon as the $C_r$ contours can be moved, all the above
$\tilde K_r$ determine in fact the same function $K(t,t')$.
So, a choice of either $\tilde K_r$ to fit the discussed $K$
is only a matter of a convenience.
The considered set of equations (56) can be also
written down in an operator form as
\begin{equation}
\tilde K=\tilde K^{(1)}+
(\hat K^{(1)}-\hat\varphi\hat f)
\tilde K
\end{equation}
where $\tilde K^{(1)}=\{\tilde K_s^{(1)}\}$ denotes the
set of the $K_s^{(1)}(t,t')$ functions, every
$K_s^{(1)}(t,t')$ being taken at $z\in C_s$.
Furthermore, $\hat\varphi=
\{\hat\varphi_{sr}\}$ is the diagonal matrix depending on
$t$. The matrix elements of this matrix are
$\hat\varphi_{ss}=\varphi_s(t)$ and $\hat\varphi_{sr}=0$
for $s\neq r$. Both matrices $\hat K^{(1)} =
\{\hat K_{sr}^{(1)}\}$ and $\hat f=\{\hat f_{sr}\}$ are
the integral operators.  For $s\neq r$, the
$\hat K_{sr}^{(1)}$ operator performs the integration
over $t'$ along $C_r$-contour, the "kernel" being
$K_{sr}^{(1)}(t,t')dt'$. We define
the "kernel" together with the differential $dt'$ to have
deal with the objects obeying bose statistics. Furthermore,
$\hat K_{ss}^{(1)}=0$.  By contrast, the kernel of
the $\hat f_{sr}$ operator is unequal to zero only for $s=r$.
And the $\hat f_{ss}$ operator performs the integration
over $t'$ along $C_b^{(s)}$-contour, its kernel being
$f_s(t')dt'$.  So, being applied to $\tilde K_r$, the
$\hat K_{sr}^{(1)}$ operator reproduces
the integral over $C_r$ in (56).
Moreover, $\hat\varphi\hat f\tilde K$ gives  the integral over
$C_b^{(s)}$ on the right side of (56).
The first term on the right side of (56)
corresponds to the first term on the right side of (58).

For both $z\in C_r$ and  $z'$ to be situated
inside $C_s$-contour ( but outside
$C_b^{(s)}$-contour ) the solution of (58) can be
written down as
\begin{eqnarray}
K_{rs}=K_{rs}^{(1)}\delta_{rs}+
\hat K_{rs}K_{ss}^{(1)}
+\frac{1}{2}\varphi_r V_{rs}^{-1}f_s-\sum_m\varphi_r
V_{rm}^{-1}\hat f_{mm}\hat K_{ms}K_{ss}^{(1)}+ \nonumber\\
\frac{1}{2}\sum_{n}\hat K_{rn}\hat\varphi_{nn}
V_{ns}^{-1}f_s-\sum_{q,m}\hat K_{rq}\hat
\varphi_{qq}
V_{qm}^{-1}\hat f_{mm}\hat K_{ms}K_{ss}^{(1)}
\end{eqnarray}
where $K_{rs}$ and $K_{rs}^{(1)}$ specifies $K(t,t')$ and
$K^{(1)}(t,t')$, respectively, provided
that $z\in C_r$, $z'$ being situated inside $ C_s$ ( but outside
$C_b^{(s)}$-contour ). Moreover, $f_s=f_s(t')$ and
$\varphi_r=\varphi_r(t)$ where $\varphi_r(t)$ and $f_s(t')$ are
defined by (49). Both
integral operator  $\hat K=\{\hat
K_{pq}\}$ and the matrix $V=\{V_{mn}\}$  are defined as
\begin{equation}
\hat K=(I-\hat K^{(1)})^{-1}-I\quad{\rm and}\quad
V_{mn}=\sum_{p}\hat f_{mp}\hat K_{pn}\varphi_n
\end{equation}
where
$I$ is the identical operator and $\hat K^{(1)}$ is the same as in
(58).  The $\{V_{mn}\}$ matrix is
defined only for those $(m,n)$, which label the odd genus-1 spin
structures. So, for superspin structures where no the genus-1 odd
superspin ones only two first terms remain on the right side
of(59).  In this case
\begin{equation}
K_{rs}=K_{rs}^{(1)}\delta_{rs}+
\hat K_{rs}K_{ss}^{(1)}.
\end{equation}
Eqs. (59) and (61) can be verified by the
substitution into eq.(58). As for eq.(53)
it can be proved that the solution of (56)
possesses  the required properties (32)-(35),
but in details this matter is planned to  discuss in an
another paper. Eqs. (59) and (61)
will be used in Sec.VI for the calculation of
the partition functions.

\section{Ghost Vacuum Correlators}
As it has been explained in Sec.III, the ghost vacuum
correlators $G_{gh}(t,t')$ satisfy eqs.(39).
For the superspin structures
where $l_{1s}=0$ for every $s$,  the above correlators
have been obtained in [11,12,30].  In this
Section we calculate the discussed correlators for those
even superspin structures where at
least  one of $l_{1s}$-characteristics is unequal to zero.

It is  convenient to write [11,12]
the discussed $G_{gh}(t,t')$  in
the terms of a new Green function G(t,t') as
\begin{equation}
G_{gh}(t,t')=G(t,t')-
\sum_{N_o,N_o'}Y_{b,N_o}(t)A_{N_oN_o'}^{-1}
\chi_{N_o'}(t')
\end{equation}
where $\chi_{N_o}$ are $3/2$-supertensors, $A_{N_o,N_o'}$
is a matrix that will be defined below and
$Y_{b,N_o}^{(s)}(t)$ are defined by (40) at $N_r=N_o$.
In contrast to eqs.(39),
the summation in (62) is performed over those
the $N_o$ indices that are associated with those $(3|2)$
Schottky parameters $q_{N_o}$, which are not moduli.
The $G(t,t')$ Green function changes under
the $\Gamma_{a,s}(l_{1s})$ and
$\Gamma_{b,s}(l_{2s}))$ mappings as follows
$$G(t_s^a,t')=Q_{\Gamma_{a,s}}^{-2}(t)\left(G^{(s)}(t,t')+
\sum_{r=1}^n\sum_{N_r}
Y_{a,N_r}^{(s)}(t)\chi_{N_r}(t')\right)\quad,$$
\begin{equation}
G(t_s^b,t')=Q_{\Gamma_{b,s}}^{-2}(t)\left(G(t,t')+
\sum_{r=1}^n\sum_{N_r} Y_{b,N_r}^{(s)}(t)
\chi_{N_s}(t')\right)
\end{equation}
where $G^{(s)}(t,t')$  has been obtained from $G(t,t')$ by
$2\pi$-twist about $\hat C_s^{(-)}$-circle (20).
Unlike (39), the summation in (63)  performs
over all the  $N_r$ indices, the $\{N_o\}$
set being among of them. The polynomials $Y_{a,N_r}^{(s)}(t)$ and
$Y_{b,N_r}^{(s)}(t)$ are defined for all $N_r$ by (40). As
it follows from (40), only the $r=s$ terms  contribute to
(63).

The $A_{N_o,N_o'}$ matrix in (62) is the square
submatrix $B_{N_oN_o'}$ of the $B_{N_sN_r}$ matrix  that is
determined by the following relations:
\begin{equation}
Y_{b,N_r}(t_s^p)=Q_{\Gamma_{p,s}}^{-2}(t)\left(Y_{b,N_r}(t)+
\sum_{N_s} Y_{p,N_s}(t) B_{N_sN_r}\right).
\end{equation}
One can calculate the discussed $B_{N_sN_r}$  matrix
using eqs. (A1) together with (A3) and
(A4).
It is quite important that $B_{N_sN_r}$ appear to be the same
for both $t\rightarrow t_s^a$ and $t\rightarrow t_s^b$ mappings.
So $G(t,t')$ being defined by(62),
satisfies  both equations  (39). It is  obvious
that the discussed $B_{N_sN_r}$ are independent of the superspin
structure. Furthermore, one can verify that
$B_{N_rk_r}=B_{k_rN_r}=0$.
Also, it can be noted that the $\tilde\chi_N$ zero modes are
calculated in the terms of $\chi_N$ as
\begin{equation}
\tilde\chi_{N_s}(t)=\chi_{N_s}(t)-
\sum_{N_o,N_o'}B_{N_sN_o}A_{N_oN_o'}^{-1} \chi_{N_o'}(t').
\end{equation}

Under the $\Gamma_{a,s}(l_{1s})$ and
$\Gamma_{b,s}(l_{2s})$ mappings  on the $t'$ supermanifold the
discussed $G(t,t')$ is due to be 3/2-superconformal  tensor.
This condition together with eqs.(63)
determines in the unique way both $G(t,t')$ and $\chi_N$.
To prove this statement we are based on the following relations:
\begin{equation}
\int\limits_{C_b^{(s)}}G(t,t')\frac{d\theta' dz'} {2\pi
i}Y_{b,N_s}(t') +\int\limits_{C_a^{(s)}}G(t,t')\frac{d\theta'
dz'} {2\pi i}Y_{a,N_s}(t') =0
\end{equation}
where  the $C_b^{(s)}$ contour is the same as in (50)
and (C10). The  integral along $C_a^{(s)}$  presents
in (66)  only for $l_{1s}\neq0$. As it has been noted in
Sec.III, in this case the cut $\tilde C_s$ appears on the
considered $z$-plane.  The one of its endcut points
is placed inside the $C_s^{(-)}$ circle and the other
endcut point is placed inside the $C_s^{(+)}$ one,
both the circles being defined by
(20).  The $C_b^{(s)}$ contour being
considered on the $z_s$ complex plane (23), starts
with the point $z_s^{(-)}$ of the intersect of
the $\hat C_s^{(-)}$ contour
with the above $\tilde C_s$ cut.  On the considered
$z_s$ complex plane (23) the $C_a^{(s)}$ path
goes along the low edge of the $\tilde C_s$ cut from the
$z_s^{(-)}$ point to the intersect
$z_s^{(+)}=\tilde C_s\bigcap \hat C_s^{(+)}$ that is
chosen to be the same as in (C10) of Appendix C.
Eqs.(66)  are proved in
the manner that is quite similar to the
proof of eqs.(50),
for more details see eq.(C14) of Appendix C.

To prove that $G(t,t')$ is determined by (63) in the
unique way, we assume that there are two different Green
functions $G_1(t,t')$ and $G_2(t,t')$  satisfying (63)
and consider the difference $\delta G(t,t')= G_1(t,t')-
G_2(t,t')$. The above $\delta G(t,t')$ has not the pole at
$z=z'$. Besides, it satisfies  eqs.(63) with some
3/2-supertensors $\delta\chi_{N_s}(t')$ instead of
$\chi_{N_s}(t')$.  Furthermore, $\delta G$ is
represented by the integral over $t''$ performed
along infinitesimal contour $C(z)$,
which surrounds $z$-point, the integrand being $\delta G(t,t'')
G_1(t'',t')$. Being  reduced
to the integrals along both $C_b^{(s)}$ and
$C_b^{(s)}$ contours, the above  integral  is found to be
equal to zero owing to  (66).  Therefore, $\delta
G(t,t')\equiv0$. It proves the desired statement that the
$G(t,t')$ Green function is
unique.  At the same time eqs.(63) determine
$\chi_{N_s}(t')$ in the unique way, too.
A like consideration
could be performed also to prove the uniqueness of
$G_{gh}(t,t')$.

If all the odd parameters are equal to zero,
$G(t,t')$  is reduced to
$G_{(o)}(t,t')$ that is given in
the terms of both the boson Green function  $G_b(z,z')$ and
the fermion Green function $G_f(t,t')$  as
\begin{equation}
G_{(o)}(t,t')=G_b(z,z')\theta'+ \theta G_f(z,z').
\end{equation}
The boson Green function $G_b(z,z')$ has been given in
[10] as
\begin{equation}
G_b(z,z')=-\sum_\Gamma\frac{1} {[z-g_\Gamma(z')][c_\Gamma
z'+d_\Gamma]^4}.
\end{equation}
In (68)  the summation is
performed over all the group products
$\Gamma=\{z\rightarrow g_\Gamma(z)=(a_\Gamma z+b_\Gamma )
(c_\Gamma z+d_\Gamma)^{-1}\}$ of the basic Schottky
group elements $\Gamma_s=\{z\rightarrow g_s(z)\}$.
As it will be explained below,
the fermion Green function $G_f(z,z')$ in eq.(67)
is calculated in the
terms of the following $G_{(\sigma)}(z,z')$ Green functions:
\begin{equation}
G_{(\sigma)} (z,z')=
\sum_\Gamma\frac{\exp\pi i
[\Omega_\Gamma(\{\sigma_s\})+\sum_s2l_{1s}
\sigma_s(J_{(o)s}(z)-J_{(o)s}(z'))]}
{[z-g_\Gamma(z')][c_\Gamma z'+d_\Gamma]^3}
\end{equation}
where $\sigma_s=\pm1$. So, $G_{(\sigma)} $ depends on a choice
of the $\{\sigma_s\}$ set.
The value $\Omega_\Gamma(\{\sigma_s\})$
in (69) is defined as
\begin{equation}
\Omega_\Gamma(\{\sigma_s\})=-\sum_{s,r}2l_{1s}
\sigma_s\omega_{sr}^{(o)}n_r(\Gamma)+
\sum_r(2l_{2r}-1)n_r(\Gamma)
\end{equation}
where $n_r(\Gamma)$ is the
number of times that the $\Gamma_r$ generators  are present
in $\Gamma$
(for its inverse $n_r(\Gamma)$ is defined to be negative ).
Furthermore, $J_{(o)s}$ in (69) are the periods of
the boson Green function $R_b(z,z')$  in
eq.(44), and $\omega_{sr}^{(o)}$ in (70) is the
period matrix at zero odd moduli.

For zero values of the odd
Schottky parameters,  $2\pi$-twist about every $B_s$-cycle is
given by   the mapping  $\{z\rightarrow z_{(o)r}=g_r(z),
\theta\rightarrow
\theta_{(o)r}=-(-1)^{l_{1r}+l_{2r}}(c_rz+d_r)^{-1}\theta)\}$.
It is follows from (69) that the changes of $\theta
G_{(\sigma)}$ under the above mappings are
\begin{equation}
\theta_{(o)r} G_{(\sigma)} (z_{(o)r},z')=(c_rz+d_r)^{-2}
\left(\theta G_{(\sigma)} (z,z')
+\sum_{N_r=\mu_r,\nu_r}\tilde Y_{\sigma,N_r}(t)
\Phi_{\sigma,{N_r}}^{(0)}(z')\right)
\end{equation}
where  $\Phi_{\sigma,{N_r}}^{(0)}(t')$  are 3/2-tensors and $\tilde
Y_{\sigma,N_r}(t)$ for $N_r=(\mu_r,\nu_r)$ is given by
\begin{equation}
\tilde
Y_{\sigma,N_r}(t)=\exp[\pi
i\sum\nolimits_s2l_{1s}\sigma_sJ_{(o)s}(z)]Y_{b,N_r}^{(0)}(t)
\end{equation}
where $Y_{b,N_r}^{(0)}(t)$ is equal to
$Y_{b,N_r}(t)$ in (40) at $\mu_r=\nu_r=0$.
One can verify that the above conditions (71) differ from
those for $G_f(z,z')$ in eq.(68). The  conditions
for $G_f(z,z')$ are derived from eqs.(63) at  zero odd
Schottky parameters.

To obtain the
$G_f(z,z')$ Green function we write $G_f(z,z')$ as the
integral over $t''$ performed along the infinitesimal contour
$C(z')$ around $z'$, the integrand being $G_{(\sigma)} (z,z'')
\theta'' G_f(z'',z')$. Running this contour away, we obtain the
following expression for $G_f(z,z')$ in question:
\begin{equation}
G_f(z,z')=
G_{(\sigma)} (z,z')-\sum_{s=1}^n\sum_{p=a,b}\sum_{N_s}
\int\limits_{C_p^{(s)}}
G_{(\sigma)} (z,z'')\frac{d\theta'' dz''} {2\pi i}
Y_{p,N_s}^{(0)}(t'')\chi_{N_s}^{(0)}(z')
\end{equation}
where $N_s=(\mu_s,\nu_s)$ and $\chi_{N_s}^{(0)}(z')$ are
3/2-tensors.  The $C_p^{(s)}$-contours ( where $p=a,b$ )
are the same as in (66).
Furthermore, $Y_{p,N_s}^{(0)}$ are
defined to be $Y_{p,N_s}$ in eq.(40) at $\mu_r=\nu_r=0$.
Eq.(73) follows from eq.(C14)  in
Appendix C.  As soon as there is the only $G_{(o)}(z,z')$
Green function,
$G_f(z,z')$ determined by eq.(73) is, in fact,
independent of $\{\sigma_s\}$.  The above
$\chi_{N_s}^{(0)}(z')$ in (73) are calculated
in the terms of $\Phi_{\sigma,N_s}^{(0)}$
in (71) from
the following relations:
\begin{equation}
\Phi_{\sigma,N_s}^{(0)}=\sum_{N_r=\mu_r,\nu_r}
\tilde M_{N_s,N_r}
(\{\sigma_q\})
\chi_{N_r}^{(0)}\quad
{\rm{where}}\quad\tilde M_{N_s,N_r}(\{\sigma_q\})=
\sum_{p=a,b}
\int\limits_{C_p^{(r)}}
\Phi_{\sigma,N_s}^{(0)}(z)\frac{d\theta dz} {2\pi i}
Y_{p,N_r}^{(0)}(t).
\end{equation}
Eqs.(74) are obtained from condition that,
being calculated from (73),
the changes under $2\pi$-twists about
$A_s,B_s$-cycles of $G_f(z,z')$
are given by (63) at zero
values of the
odd Schottky parameters. So, the desired $G_{(o)}(t,t')$
Green function is determined by (68),
(73) and (74)

In the presence of the odd Schottky
parameters we calculate  $G(t,t')$
in the form of series over the above odd parameters,
the zero power term being $G_{(o)}(t,t')$.  For
this aim we derive the equations similar to (56)
for $K(t,t')$. For the superspin
structures without
the genus-1 odd superspin ones the above equations could be
obtained directly for the desired $G(t,t')$ in the
terms of $G_s^{(1)}(t,t')$ presenting the $G(t,t')$
Green functions associated with the genus-1  supermanifolds.
But for the odd genus-1 (super)spin structures the above
$G_s^{(1)}(t,t'$) do not exist.
In this case we can use ancillary Green functions
$S_\sigma(t,t')$ defined below. In Sec.VI we shall
see that these ancillary Green functions appears to be
useful for calculation also those superspin structures
that do not contain the odd genus-1 superspin ones.

The above genus-n Green functions
$S_\sigma(t,t')$  satisfy  the following conditions:
\begin{equation}
S_\sigma(t_s^b,t')=Q_{\Gamma_{b,s}}^{-2}(t)
\left(S_\sigma(t,t')+\sum_{N_s}
\hat Y_{\sigma,N_s}^{(1)}(t)\Psi_{\sigma,N_s}(t')\right)
\end{equation}
where $N_s=(k_s,u_s,v_s,\mu_s,\nu_s)$ and
$\Psi_{\sigma,N_s}(t')$ are 3/2-supertensors. The $\hat
Y_{b,N_s}^{(1)}(t)$ polynomials for $N_s=(k_s,u_s,v_s)$
are equal to $Y_{b,N_s}^{(0)}(t_s)
Q_{\tilde\Gamma_s}^{-2}(t_s)$
where $Y_{b,N_s}^{(0)}(t)$ are defined by
eq.(40) at $\mu_s=\nu_s=0$.  For $N_s=(\mu_s,\nu_s)$,
the discussed $\hat
Y_{\sigma,N_s}(t)$ polynomials are equal to $\tilde
Y_{p,N_r}(t_s)Q_{\tilde\Gamma_s}^{-2}(t_s)$,
$\tilde Y_{\sigma,N_r}(t)$ being defined by (72)
for the genus $n=1$.  The $t_s$ transformation is
defined by (23).  In addition to (75), the
$S_\sigma(t,t')$ Green function is defined to have no periods
about $A_s$ cycles.

The desired $G(t,t')$ in eq.(62) is calculated
in the terms of $S_\sigma(t,t')$ as follows
\begin{equation}
G(t,t')=S_\sigma(t,t')-\sum_{s=1}^n\sum_{p=a,b}\sum_{N_s}
\int\limits_{C_p^{(s)}}S_\sigma(t,t_1) \frac{d\theta_1
dz_1} {2\pi i}Y_{p,N_s}(t_1)\chi_{N_s}(t')
\end{equation}
where $\chi_{N_s}(t')$ are determine by
\begin{equation}
\Psi_{\sigma,N_s}=\sum_{N_r}
M_{N_s,N_r}(\{\sigma_q\})\chi_{N_r}\quad
{\rm{where}}\quad  M_{N_s,N_r}(\{\sigma_q\})=
\sum_{p=a,b}
\int\limits_{C_p^{(r)}}
\Psi_{\sigma,N_s}(z)\frac{d\theta dz} {2\pi i}
Y_{p,N_r}(t).
\end{equation}
Eq.(76) is obtained by the same method
that has been used to derive eq.(73).
Eqs.(77) are obtained
from condition that the changes of $G(t,t')$
under $2\pi$-twists about $A_s,B_s$-cycles are given by
(63). Being unique, as it has been proved above,
$G(t,t')$ determined by eq.(73) is, in fact,
independent of $\{\sigma_s\}$.

If all the odd Schottky parameters are equal to zero, then
$S_\sigma(t,t')$ is reduced to $S_{(o)\sigma}$.
The above $S_{(o)\sigma}$ has the following
form:
\begin{equation}
S_{{(o)}\sigma}(t,t')=G_b(z,z')\theta'+\theta
S_{(f)\sigma}(z,z')
\end{equation}
where $G_b(z,z')$ is given by (68) and
$S_{(f)\sigma}(z,z')$ are  calculated in the
terms of the above $G_{(\sigma)} (z,z')$ Green function
(69), as
\begin{equation}
S_{(f)\sigma}(z,z')=G_{(\sigma)}
(z,z')-\sum_{r=1}^n\sum_{N_r=\mu_r,\nu_r}
\int\limits_{C_b^{(r)}}
G_{(\sigma)} (z,z'')\frac{d\theta'' dz''} {2\pi i}
\tilde Y_{\sigma,N_r}^{(1)}(t'')
\Psi_{\sigma,N_r}^{(0)}(z')
\end{equation}
where $\tilde Y_{\sigma,N_r}^{(1)}(t'')$ is equal to
$\tilde Y_{\sigma,N_r}(t'')$ defined by
eq.(72) at the genus $n=1$, and $\Psi_{N_r}^{(0)}(z')$
is equal to
$\Psi_{N_s}(t')$ in eq.(75) at zero values of
the odd Schottky parameters. The above
$\Psi_{\sigma,N_r}^{(0)}(z')$ calculated in the terms of
$\Phi_{\sigma,N_s}^{(0)}$ in (71) from the following
relations:
\begin{equation}
\Phi_{\sigma,N_s}^{(0)}=\sum_{N_r=\mu_r,\nu_r}\hat
M_{N_s,N_R}(\{\sigma_p\}) \Psi_{\sigma,N_r}^{(0)}\quad
{\rm{where}}\quad \hat M_{N_s,N_R}(\{\sigma_p\})=
\int\limits_{C_b^{(r)}}
\Phi_{\sigma,N_s}^{(0)}(z)\frac{d\theta dz} {2\pi i}
\tilde Y_{\sigma,N_r}^{(1)}(t).
\end{equation}
Eqs.(80) are kindred
eqs.(74). They
are obtained from condition that
the changes under $2\pi$-twists about $A_s,B_s$-cycles
of the $S_{(f)\sigma}(z,z')$ Green functions in (79)
are given by (75) at zero values of the odd
Schottky parameters.

At arbitrary odd Schottky parameters the $S_\sigma(t,t')$ Green
function can be calculated the equations
kindred to eqs.(56) of the previous section.
These equations give $S_\sigma(t,t')$ in the
form of series over odd parameters, the zero power term being
$S_{(o)\sigma}$.  To obtain the above equations we   use
the genus-1 Green functions
$S_{\sigma,s}^{(1)}(t,t')$ defined by
\begin{equation}
S_{\sigma,s}^{(1)}(t,t')=Q_{\tilde\Gamma_s}(t_s)^{-2}\left
[G_b^{(1)}(z_s,z_s')
\theta_s'+\theta_sG_{(\sigma)}^{(1)}(z_s,z_s')-
\tilde\varepsilon_s'\Sigma_\sigma(z_s')\right]
Q_{\tilde\Gamma_s}(t_s')^3
\end{equation}
where $t_s$  is defined by (23)
and the $Q_{\tilde\Gamma_s}$ factor is defined by (28).
Apart
the last term, $S_{\sigma,s}^{(1)}(t,t')$  is obtained by
the $\tilde\Gamma_s$ transformation (23) of
$G_b^{(1)}(z,z')\theta'+\theta G_{(\sigma)}^{(1)}(z,z')$.
The last term being 3/2-supertensor, is determined from
the condition that $S_{\sigma,s}^{(1)}(t,t')$ decreases
at $z\rightarrow\infty$ or
$z'\rightarrow\infty$.   It is follows from (81)
that $S_{\sigma,s}^{(1)}(t,t')$ has no periods about $A_s$
cycles. And under the $\Gamma_{b,s}$ transformation
$\{t\rightarrow t_s^b\}$ the discussed
$S_{\sigma,s}^{(1)}(t,t')$ is changed as
\begin{equation}
S_{\sigma,s}^{(1)}(t_s^b,t')=Q_{\Gamma_{b,s}}^{-2}(t)
\left(S_{\sigma,s}^{(1)}(t,t')+\sum_{N_s}
\hat Y_{\sigma,N_s}^{(1)}(t)
\Psi_{\sigma,N_s}^{(1)}(t')\right)
\end{equation}
where $N_s=(k_s,u_s,v_s,\mu_s,\nu_s)$. Furthermore, $\hat
Y_{b,N_s}^{(1)}(t)$ are the same as in (75) and
$\Psi_{N_s}^{(1)}(t')$ are 3/2-supertensors.

The desired set of equations for $S_\sigma(t,t')$ is given by
\begin{equation}
S_\sigma(t,t')=S_{\sigma,s}^{(1)}(t,t')+\sum_{r\neq
s}\int\limits_{C_r}S_{\sigma,s}^{(1)}(t,t_{(1)})
\frac{d\theta_{(1)} dz_{(1)}} {2\pi i}S_\sigma(t_1,t')\quad{\rm
for}\quad s=1,2,...n
\end{equation}
Eqs.(83) are obtained by the method employed above to
derive eqs.(56).
There is no the term with $r=s$ in the sum over
$r$.  Indeed, as far
as the $\Gamma_{a,s}$ and $\Gamma_{b,s}$
mappings are the same for
both $S_\sigma$ and $S_{\sigma,s}^{(1)}$, the discussed term
appears to be equal to zero
owing to both (75) and (C15) taken for the genus
$n=1$.

Every  $ C_r$-contour being considered, we denote
$S_\sigma(t,t')$  for $z\in C_r$  as
$\tilde S_{\sigma,r}(t,t')$. We
use (83)  to calculate the set $\{\tilde S_\sigma\}$ of
the above $\tilde S_{\sigma,r}$  functions. This procedure is
quite similar to that discussed in the previous Section.  The
discussed set of equations can  be  written down in the
operator form similar to (58) as
\begin{equation}
\tilde
S_\sigma=\tilde S_\sigma^{(1)}+
\hat S_\sigma^{(1)}\tilde
S_\sigma
\end{equation}
where $\tilde
S_\sigma^{(1)}$ is the set of the
$S_{\sigma,s}^{(1)}(t,t')$
functions,  matrices $\hat S_\sigma^{(1)} =\{\hat
S_{\sigma,sr}^{(1)}\}$  being the integral operators.
For $s\neq r$, the $\hat S_{\sigma,sr}^{(1)}$ operator
performs the integration over $t'$ along $C_r$-contour,
the kernel being $S_{\sigma,sr}^{(1)}(t,t')$.
And $\hat S_{\sigma,ss}^{(1)}=0$.

For both $z\in C_r$ and  $z'$ to be situated
inside $C_s$-contour,  the solution of (84) can be
written down as
\begin{equation}
S_{\sigma,rs}=S_{\sigma,rs}^{(1)}\delta_{rs}+
\sum_{n}\hat S_{\sigma,rn}^{(1)}(I-\hat
S_\sigma^{(1)})_{ns}^{-1} S_{\sigma,ss}^{(1)}
\end{equation}
where $S_{\sigma,rs}$ and $S_{\sigma,rs}^{(1)}$ specifies
$S_\sigma(t,t')$ and $S_\sigma^{(1)}(t,t')$,
respectively, provided
that $z\in C_r$, $z'$ being situated inside $C_s$.
Eq.(85) can be verified by the substitution
into eq.(84).
It can be proved that (85)
possesses  the required properties (63).
It follows from (85) that the
3/2-forms $\Psi_{\sigma,N_s}(t')$ provided that $z'$
is situated inside $C_s$ can be written down as
\begin{equation}
\Psi_{\sigma,N_s}=\Psi_{\sigma,N_s}^{(1)}+
\sum_{n\neq s}\hat\Psi_{\sigma,N_sn} ^{(1)}(I-\hat
S_\sigma^{(1)})_{ns}^{-1} S_{\sigma,ss}^{(1)}
\end{equation}
where $\Psi_{\sigma,N_sn}^{(1)}$ is the integral operator that
performs the
integration over $t_1$ along $C_n$-contour, its kernel being
$\Psi_{\sigma,N_s}^{(1)}(t_1)dt_1$. In the next Section
we use (85) and (86) to determine the partition
functions in (15).

\section{Calculation of the partition functions}
Having the vacuum correlators to be known, in this Section we
calculate from eq.(41) the partition functions $\hat
Z_{L,L'}^{(n)}$ in eq.(15).
We continue this
calculation in the next Section VII where the
final formulae for the partition functions will be given.

To solve eq.(41) it is useful to remove the
dependence on $t$ in (41) employing for this aim the
following relations
\begin{equation}
\sum_{p=a,b}\int\limits_{C_p^{(r)}}\tilde
\chi_{N_s}(t)\frac{d\theta
dz} {2\pi i}Y_{b,N_s}(t)=\delta_{N_sN_r}\quad{\rm for}\quad
(N_r,N_s)\in\{N\}.
\end{equation}
In (87) the $\{N\}$ set is associated with those
Schottky parameters, which are  moduli. To obtain
(87) we represent $\tilde\chi_{N_s}(t)$
in the form of the integrals along the infinitesimal
contour $C(z)$ surrounding $z$-point.  The above contour
is transformed to obtain the sum of the integrals along
the $C_r$ contours that reduced to
(87). For this aim we use eq.(C14) of Appendix
C. Below we also will use the kindred relations for
both $\chi_{N_s}(t)$ in (63) and
$\Psi_{\sigma,N_s}(t)$ in (75). The
above relations are given by
\begin{equation}
\sum_{p=a,b}\int\limits_{C_p^{(r)}}
\chi_{N_s}(t)\frac{d\theta
dz} {2\pi i}Y_{b,N_s}(t)=\delta_{N_sN_r}
\quad{\rm and}\quad
\int\limits_{C_b^{(r)}}\frac{d\theta dz} {2\pi i}
\Psi_{\sigma,N_s}(t)\hat Y_{\sigma,N_s}^{(1)}(t)=
\delta_{N_s,N_r},
\end{equation}
see Appendix C for more details.
Unlike (87), (88)
are true for all $N_s$ including the $\{N_o\}$ set, as well.

By means of (87), we
transform (41) to the set of equations
\begin{equation}
\partial
q_{N_r}\ln{\hat
Z_{L,L'}^{(n)}}=
\sum_{p=a,b}\int\limits_{C_p^{(r)}}\left[<T_{gh}(t)>
+<T_m(t)>- \sum_N \frac{\partial}{\partial q_N}
\tilde\chi_N(t)\right]dtY_{p,N_r}(t)
\end{equation}
where $dt=d\theta dz /2\pi i$. The summation over
$N$ in (89) is performed as in (41).
As in (41),
the derivatives with respect to odd moduli in eq.(89)
are implied to be  the "right" ones.

Both $<T_{gh}(t)>$ and $<T_m(t)>$ in (89)
are calculated from (42) and (43) in the
terms of the $G_{gh}(t,t')$ and $K(t,t')$
Green functions given in the previous Sections.

Eqs.(89) determine the partition functions up to an
arbitrary factor  independent of the moduli. But this  factor
may depend on $n$, $L$ and  $L'$, as well. In addition, the
discussed factor may depend on the $\{q_{N_o}\}$ set of those
$(3|2)$ Schottky parameters, which are chosen to be the same for
all the genus-$n$ supermanifolds and, therefore, they
are not moduli. Below we choose the above
$\{q_{N_o}\}$ set as $\{q_{N_o}\}=
(u_1,v_1,u_2|\mu_1,\nu_1)$.

The dependence on the above $(3|2)$ Schottky
parameters determined from the condition
that the superstring amplitudes are independent of a choice of
these $(3|2)$ parameters [11,12].
Moreover, the dependence on $n$, $L$ and $L'$ in the discussed
factor is calculated from the condition of the supermodular
invariance together with the factorization requirement when the
handles move away from each other. So, only a coupling
constant remains to be arbitrary. We write
down the $\hat Z_{L,L'}^{(n)}$ partition functions as
follows
\begin{equation}
\hat Z_{L,L'}^{(n)}=g^{2n}\left[\det2\pi
i[\overline{\omega(\{q_{N_s}\},L')}-
\omega(\{q_{N_s}\},L)]\right]^{-5}
Z_L^{(n)}(\{q_{N_s}\})
\overline {Z_{L'}^{(n)}(\{q_{N_s}\})}
\end{equation}
where $g$ is a coupling constant,  $\omega(\{q_{N_s}\},L)$ is
the period matrix and  $Z_L^{(n)}(\{q_{N_s}\})$ is
the holomorphic function of the  $\{q_{N_s}\}$
Schottky parameters. This
holomorphic structure of $\hat Z_{L,L'}^{(n)}$
arises because the superstring modes can be divided into
left and right movers.

The only non-holomorphic factor on the right side of (90)
originates from the proportional to
$[\overline{\omega(\{q_{N_s}\},L')}-\omega(\{q_{N_s}\},L)]^{-1}$
terms in the scalar superfield vacuum correlator (31). The
contribution of these terms into the right side of
(89) is
\begin{equation}
5[2\pi i(\overline{\omega(\{q_{N_s}\},L')}-
\omega(\{q_{N_s}\},L))^{-1}]_{mn}
\sum_{p=a,b}\int\limits_{C_p^{(r)}}2\pi iD(t)\eta_m(t;L)
2\pi i\eta_n(t;L)
dtY_{p,N_r}(t)
\end{equation}
where $\eta_n(t;L)$ is the half-form (37).
To  obtain the discussed non-holomorphic factor in (90)
we prove firstly that the integral in (91) is equal to
$2\pi i \partial_{N_r}\omega_{mn}(\{q_{N_s}\},L)$. This proof
employs the transformation low under the
$\Gamma_{p,s}$ mappings ( $p=a,b$ ) of $\partial_{N_s}F_q(t)$,
which can be done provided that the
above transformation low for $F_q(t)$ is known. To obtain the
desired transformation low one is due only to express the
$\partial_{N_s}F_q(\Gamma_{p,s}(t))$ derivatives taken for
$\Gamma_{p,s}(t)$ to be constant in the terms of
those calculated for $t$ being constant. If
$F_p\rightarrow Q_{\Gamma_{p,s}}^q\tilde F_q$ under the
considered  mapping, then we obtain for the
derivative that
\begin{equation}
\partial_{N_s}F_q(t)\rightarrow
Q_{\Gamma_{p,s}}^q(t)\left[\partial_{N_s}\tilde F_q(t)-
E_{p,N_s}^{(q)}\left(\tilde F_q(t)\right)\right]
\end{equation}
where $E_{p,N_s}^{(q)}( F(t)$ is defined by
\begin{equation}
E_{p,N_s}^{(q)}(F(t))=\frac{q}{2}F(t)\partial_zY_{p,N_s}(t)+
\frac{\epsilon(F)}{2}[D F(t)]DY_{p,N_s}(t)+
[\partial_zF(t)]Y_{p,N_s}(t).
\end{equation}
In this equation $D$ is the spinor derivative (27).
Furthermore, $\epsilon(F)=1$, if $F_q$ obeys the fermion statistics
and $\epsilon(F)=-1$ otherwise. To obtain the above formulae we
use  for the derivatives of $Y_{p,N_s}(t)$ eqs.(A6) of
Appendix A.

Employing (92), in Appendix D we  derive  the
representation of $\partial_{N_r} J_m(t)$ in the form of the
integral along $C_a^{(r)}$ and $C_b^{(r)}$. The desired
statement about the integral in (91) follows from
this representation for $\partial_{N_r} J_m(t)$, as
it is shown  in Appendix D.

As far as the integral in (91) is found to be
$2\pi i \partial_{N_r}\omega_{mn}(\{q_{N_s}\},L)$, one
concludes that (91) present none other than
the derivative with respect to  $N_r$ of the non-holomorphic
factor in (90). The above non-holomorphic factor  already
discussed  in [23,30].
This non-holomorphic factor  extends
to depending on $L$ period matrices the well known
non-holomorphic multiplier [7] arising in the boson string
theory.

To determine the holomorphic $Z_L^{(n)}(\{q_{N_s}\})$
factor in (89), it is convenient to integrate  by
parts the $\partial (FB)- D[(DF)B]/2$ terms in $<T_{gh}>$.
Simultaneously we carry out
the $\partial q_N$ with $\tilde\chi_{N_s}$
to $Y_{p,N_s}(t)$. For this aim we  take into account
eqs.(87).  The above procedure,
as well as the integration
by parts originates the out integral terms.  These
terms disappear owing to (A5)  with
the exception of those $I_{(an)}$ terms, which are due to
the conformal anomaly in $(FB)$
and in $(DF)B$.  These $I_{(an)}$ terms will
be canceled by kindred terms appearing in the following
calculations.  The resulting
equations for $Z_L^{(n)}(\{q_{N_s}\})$ turn out to be
\begin{equation}
\partial_{q_{N_r}}\ln
Z_L^{(n)}(\{q_{N_s}\})= \sum_{p=a,b}
\int\limits_{C_p^{(r)}}dt
W_{p,N_r}(t)+ \sum_{p=a,b}\int\limits_{C_p^{(r)}}dt
W_{p,N_r}^{(o)}(t)+I_{(an)}
\end{equation}
where $I_{(an)}$ are
the out integral terms discussed above.
Furthermore,
$W_{p,N_r}(t)$  in  (94) is defined as
\begin{equation}
W_{p,N_r}(t)=-5\partial_z K(t,t)Y_{p,N_r}+
E_{p,N_r}^{(-2)}\left( G(t,t)\right)- \sum_N
(-1)^{e(N)e(N_r)} \chi_N(t)\partial q_NY_{p,N_r}(t)
\end{equation}
where $E_{p,N_r}^{(-2)}$ is defined by (93) at
$q=-2$ and the summation over $N$ performs as in
(89). Moreover, $e(N_r)=1$ for
$N_r=(\mu_r,\nu_r)$ and $e(N_r)=0$ for $N_r=(k_r,u_r,v_r)$.
In the accordance with the general prescription [25]
both the first term and the second one  in (95)
are defined as the limit at $t'\rightarrow t$ of
$-5\partial_z K(t,t')Y_{p,N_r}+
E_{p,N_r}^{(-2)}\left( G(t,t')\right)$,
the singular term $(\theta-\theta')(z-z')^{-1}$ being
omitted in both $K(t,t')$ and  $G(t,t')$. The Green function
$K(t,t')$ appears in (95) instead of the vacuum
correlator (31) of the scalar superfields because
the difference of above (31) and
$K(t,t')$ has been already taken into account by the above
discussed non-holomorphic factor in (90).  Besides, in
(95) we use $G(t,t')$ instead of the ghost correlator
$G_{gh}(t,t')$. The difference of $G_{gh}(t,t')$ and $G(t,t')$
being calculated from (62), contributes into
(94) by the second term on the right side of
(94) with $ W_{p,N_r}^{(o)}(t)$ to be
\begin{equation}
W_{p,N_r}^{(o)}(t)=(-1)^{e(N_r)}\sum_{N_o,N_o'}\left[
E_{p,N_o}^{(-2)}\left(Y_{b,N_r}(t)\right)+
\sum_N [\partial q_NY_{p,N_r}(t)]
B_{NN_o}\right]A_{N_oN_o'}^{-1}
\chi_{N_o'}(t)
\end{equation}
where the $B_{N_sN_o}$ matrix  is determined in (64)
amd the $A_{N_oN_o'}$ matrix in
(94) is the same as in (62).

In the following consideration we express
$G(t,t')$  in (95) in the terms of
$S_\sigma(t,t')$. For this aim we employ eq.(76).
The integrals over $t_1=(z_1|\theta_1)$ in (76) give
raise the integrals in (94) over both $t$ and $t_1$.
In these integrals the integration over $t_1$ is performed
before the integration over $t$, but it will be
convenient for the solution of eqs.(94) to change
the above order of the integration
and to perform the integration
over $t$ before the integration over $t_1$. In this
case one is due to be careful with the terms
associated with  $s=r$ in the sum over $s$ in (76).
Indeed, the above terms contribute into (89) as
\begin{equation}
-\sum_{p=a,b}\sum_{N_r'}
\int\limits_{C_p^{(r)}}(-1)^{e(N_r)e(N_r')+e(N_r)}
dt\chi_{N_r'}(t)E_{p,N_r}^{(-2)}
\left(S_\sigma(t,t_1)
\right)dt_1Y_{p,N_r'}(t_1)
\end{equation}
where the integrations over both $t$ and $t_1$ are performed
along the same path. So, changing the order of the
integration over $t$ and $t_1$, one meets with ambiguities due
to the pole at $z=z_1$ in $S_\sigma(t,t_1)$. To avoid these
ambiguities, we note
that in (97) the terms with $N_r=k_r$ are absent  because
of eqs.(C15) and also because $Y_{b,k_s}(t)=\tilde
Y_{\sigma,k_s}^{(1)}(t)$ and $Y_{a,k_s}(t)=0$, as well.
Therefore, in (97) the integrals over $t_1$ can be
given in the form of the integrals over the $C_r$ contour
as follows
\begin{equation}
-\sum_{N_r''}
\int\limits_{C_r}S_\sigma(t,t_1)dt_1
Y_{p,N_r''}(t_1)A_{N_r''N_r'}^{-1}
\end{equation}
where the $C_r$ contour is defined in Sec.IV and Sec.V
and $A_{N_r''N_r'}$ is the submatrix
$B_{N_r''N_r'}$ of the $B_{N_sN_r}$ matrix (64).
In (98) there is implied that the pole at$z_1=z$ is
situated outside of the $C_r$ contour on the $z_1$ complex
plane. But we  accommodate the above pole inside the $C_r$
contour. At the same time, we add in (98) the suitable
terms to cancel the appearing contribution of the discussed
pole.  In this case the desired change of the order of the
integration in (97) can be performed without doubt,
the result being
\begin{eqnarray}
\sum_{N_r',N_r''}\sum_{p=a,b}
\int\limits_{C_r}dt_1Y_{p,N_r''}(t_1)A_{N_r''N_r'}^{-1}
\int\limits_{C_p^{(r)}}\chi_{N_r'}(t)dt
E_{p,N_r}^{(-2)}(S_\sigma(t,t_1))- \nonumber \\
\sum_{N_r',N_r''}\sum_{p=a,b}\int\limits_{C_p^{(r)}}
E_{p,N_r}^{(-2)}(Y_{p,N_r''}(t))A_{N_r''N_r'}^{-1}
\chi_{N_r'}(t)dt .
\end{eqnarray}
After the  $(t_1\rightarrow t,t\rightarrow t_1)$
redefinition in   (99),
eqs.(94) turn out to be
\begin{equation}
\partial_{q_{N_r}}\ln Z_L^{(n)}(\{q_{N_s}\})=
\sum_{p=a,b}\int\limits_{C_p^{(r)}}dt
W_{p,N_r}^{(o)}(t)+
\sum_{p=a,b}\int\limits_{C_p^{(r)}}dt \tilde W_{p,N_r}(t)+
\int\limits_{C_r}dt \hat W_{N_r}(t)+I_{(an)}
\end{equation}
where $W_{p,N_r}^{(o)}(t)$ is defined
by (96) and $\tilde
W_{p,N_r}(t)$ is given by
\begin{eqnarray}
\tilde W_{p,N_r}(t)=-5\partial_z
K(t,t)Y_{p,N_r}+ E_{p,N_r}^{(-2)}\left(
S_\sigma(t,t)\right)- \sum_N (-1)^{e(N)e(N_r)}
\chi_N(t)\partial q_NY_{p,N_r}(t)- \nonumber \\
\sum_{N_r',N_r''}
(-1)^{e(N_r)e(N_r')+e(N_r')}
\chi_{N_r'}(t)E_{p,N_r}^{(-2)}(Y_{p,N_r''}(t))
A_{N_r''N_r'}^{-1}.
\end{eqnarray}
In (101) both the first term and the second one
are again defined as the limit at $t'\rightarrow t$
of $-5\partial_z
K(t,t')Y_{p,N_r}+ E_{p,N_r}^{(-2)}\left(
S_\sigma(t,t')\right)$,
the singular term $(\theta-\theta')(z-z')^{-1}$ being
omitted in both $K(t,t')$ and  $S_\sigma(t,t')$.
Furthermore, $\hat W_{N_r}(t)$ in (100)
is given by
\begin{equation}
\hat W_{N_r}(t)=\sum_{N_r',N_r''}
Y_{p,N_r'}(t_1)A_{N_r''N_r'}^{-1}
\sum_{p=a,b}\int\limits_{C_p^{(r)}}
\chi_{N_r'}(t_1)dt_1
E_{p,N_r}^{(-2)}(S_\sigma(t_1,t)).
\end{equation}
Eqs. (101) and (102) follow directly from
(96) and (99).

One can verifies that in the expression contained in
the big square brackets in (96)
all those terms disappear, which are more higher
degree in $(z,\theta)$, than degree-2. So
the discussed expression is the degree-2 polynomial in
$(z,\theta)$. Therefore, the calculation of the first term
in (100) can be performed by using of the integral
relations (88) for $\chi_{N_r}(t)$.  In particulary,
one obtains that in the $N_r=k_r$ case
the discussed term in (100) is equal to zero.
Indeed, one can check that in the $N_r=k_r$ case the
functions in the big square brackets in (96)
have no periods under both $\Gamma_{a,r}$ and
$\Gamma_{b,r}$ mappings. Therefore, being the degree-2
polynomial, the above functions are proportional to
$Y_{b,k_r}(t)$.
It is because only  $Y_{b,k_r}(t)$ among of the
degree-2 polynomials has no periods under twists about
$(A_r,B_r)$-cycles.  So the discussed term
in (100) is expressed in the terms of the
integrals  given in (88).
And, as far as $k_r$ are not contained in
the $\{q_{N_o}\}$ set, one concludes that the considered
term in (100) is equal to zero.

For $N_r\neq k_r$, one can verify that
the discussed first term in (100) can be given
as follows
\begin{equation}
\sum_{N_r''}\int\limits_{C_r}dt
W_{p,N_r''}^{(o)}(t)A_{N_r''N_r}^{-1}
\end{equation}
where the $C_r$ contour defined in Sec.IV and Sec.V,  the
$A_{N_r''N_r}$ matrix being the same  as in (98)
and $W_{p,N_r''}^{(o)}(t)$ is given by (96). Using
(A3), we can present the degree-2 polynomials in the
square brackets in (96) as the sum
of the $Y_{bN_r}(t)$ polynomials. We use (64) to
reduce the discussed integrals along the $C_r$ contour to
the integrals given in (88). The above reduction
can be performed because the $B$
matrix in (64)] is independent on $p$.
So, the discussed first term in (100) are
calculated without employing  the
explicit form of $\chi_{N_o}(t)$.  The result is that
the discussed first term on the right side of
(100) originates
in $Z_L^{(n)}(\{q_{N_s}\})$ only the $H_o(\{q_{N_o}\},\mu_2)$
factor as follows
\begin{equation}
H_o(\{q_{N_o}\},\mu_2)=1-\frac{\mu_1\mu_2}{2(u_1-u_2)}-
\frac{\nu_1\mu_2}{2(v_1-u_2)}.
\end{equation}
The above $H_o(\{q_{N_o}\},\mu_2)$ factor  depends on
the choice of
the $\{q_{N_o}\}$ set of parameters  that are not moduli.
Eq.(104) is obtained provided that
the  $\{q_{N_o}\}$ set
is chosen to be $(u_1,v_1,u_2|\mu_1,\nu_1)$.

To calculate the other terms on the right side
of (100) we use for $K(t,t')$ and for
$S_\sigma(t,t')$ eqs. (59) and
(85), respectively. In the following,
it will be convenient
to divide $\tilde W_{p,N_r}(t)$ into the sum of three terms as
\begin{equation}
\tilde W_{p,N_r}(t)=W_{p,N_r}^{(r)}(t)-
5\partial_z [K(t,t)-K_s^{(1)}(t,t)]Y_{p,N_r}+ W_{p,N_r}'(t)
\end{equation}
where $K_s^{(1)}(t,t)$ is defined in the terms of
$R_s^{(1)}(t,t)$ by (36). In turn, $R_s^{(1)}(t,t)$
is defined by (46). In  the
$\partial_z [K(t,t)-K_s^{(1)}(t,t)]$ term, the derivative
are taken with respect to only the first argument of both
$K(t,t)$ and $K_s^{(1)}(t,t)$.
Furthermore, $W_{p,N_r}^{(r)}(t)$ is
given by
\begin{eqnarray}
W_{p,N_r}^{(r)}(t)= \sum_{N_r'}
\Psi_{\sigma,N_r'}^{(1)}(t)
\left[E_{p,N_r}^{(-2)}\left(Y_{\sigma,N_r'}^{(1)}(t)\right)-
\partial_{q_{N_r}}Y_{p,N_r'}(t)\right]- \nonumber \\
5\partial_z K_r^{(1)}(t,t)Y_{p,N_r}+
E_{p,N_r}^{(-2)}(S_{\sigma,r}^{(1)}(t,t))
\end{eqnarray}
where $Y_{\sigma,N_r'}^{(1)}(t)$ is the same
as in (75). Moreover, $S_{\sigma,r}^{(1)}(t,t)$
and$\Psi_{\sigma,N_r'}^{(1)}(t)$ are defined by (81) and
(82). In (106) the
$-5\partial_z K_r^{(1)}(t,t)Y_{p,N_r}+
E_{p,N_r}^{(-2)}(S_{\sigma,r}^{(1)}(t,t))$ term
is again defined as the limit at $t'\rightarrow t$
of $-5\partial_z K_r^{(1)}(t,t')Y_{p,N_r}+
E_{p,N_r}^{(-2)}(S_{\sigma,r}^{(1)}(t,t'))$,
the singular term $(\theta-\theta')(z-z')^{-1}$ being
omitted in both $K_r^{(1)}(t,t')$ and
$S_{\sigma,r}^{(1)}(t,t')$.

Furthermore, $W_{p,N_r}'(t)$ in (105) is
the rest of $\tilde W_{p,N_r}(t)$ that remains after
the subtraction
of two first terms on the right side of(105). So
\begin{equation}
W_{p,N_r}'(t)=\tilde W_{p,N_r}(t)-
W_{p,N_r}^{(r)}(t)+ 5\partial_z
[K(t,t)-K_r^{(1)}(t,t)]Y_{p,N_r}
\end{equation}
where $\tilde W_{p,N_r}(t)$ is given by (101)
and the $5\partial_z [K(t,t)-K_r^{(1)}(t,t)]Y_{p,N_r}$
term is defined as in  (105).

The proportional to $\Psi_{\sigma,N_r'}^{(1)}(t)$
terms are added into (106) and, at the same time,
they are subtracted from $W_{p,N_r}'(t)$.
So $\tilde W_{p,N_r}(t)$ remains unchanged. In is
case the sum of the $W_{p,N_r}^{(r)}(t)$ contribution to
(100) and of the $I_{an}$ term  has the form
of the derivative with respect to $N_r$
of a function of the moduli.
So the considered terms in (100) originate in
$Z_L^{(n)}(\{q_{N_s}\})$ some
multiplier $\tilde Z^{(1)}(q_{N_r};l_{1r},l_{2r})$.
Simultaneously, $W_{p,N_r}'(t)$ together with
$\hat W_{p,N_r}(t)$ originates
in $Z_L^{(n)}(\{q_{N_s}\})$ the factor
$\tilde Z_{gh}^{(n)}(\{q_{N_s}\},L)$ that satisfies
the following equation
\begin{equation}
\partial_{q_{N_r}}\ln \tilde Z_{gh}^{(n)}(\{q_{N_s}\},L)=
\sum_{p=a,b}\int\limits_{C_p^{(r)}}dt  W_{p,N_r}'(t)+
\int\limits_{C_r}dt \hat W_{p,N_r}(t)
\end{equation}
where $\hat W_{p,N_r}(t)$ is defined by (102).

In the calculation of the
$\tilde Z^{(1)}(q_{N_r};l_{1r},l_{2r})$
factor that is due to (106), we consider
separately those of the proportional to
$\Psi_{\sigma,N_r'}^{(1)}(t)$ terms, which
do not contain the
derivatives of the $J_{(o)r}^{(1)}(z_r)$ functions. The above
derivatives appear, in general,  because  both
$Y_{\sigma,\mu_r}^{(1)}(t)$ and $Y_{\sigma,\nu_r}^{(1)}(t)$
may be proportional to $\exp[\pm J_{(o)r}^{(1)}]$. In the
discussed terms the polynomial in $(z,\theta)$ factors
appear to be not  higher degree,  than 2,
just as in (96). We again can use (A3)
to represent the above polynomial
factors as a sum over $\hat Y_{\sigma,N_s}^{(1)}(t)$. So the
contribution of the considered terms into (100) can be
calculated by using of relations (88) without
the knowledge of the explicit form of
$\Psi_{\sigma,N_r'}^{(1)}(t)$.

To calculate the contribution into  (100) of
the remainder of the proportional to
$\Psi_{\sigma,N_r'}^{(1)}(t)$ terms,
one needs  the explicit form of the sum over $N_r'$ of
$Y_{\sigma,N_r'}^{(1)}(t)\Psi_{\sigma,N_r'}^{(1)}(t)$.
This value can be found from (82).
It is convenient to calculate the considered terms
together with two last terms on the right side of
(106). Moreover,  it is useful
to employ the $t_s$ supercoordinates (23)
instead of $t$. Under this substitution the additional
terms arise that are  due to
the conformal anomaly in $(FB)$ and in $(DF)B$.  These
terms cancel
the $I_{(an)}$ term in (100). The desired
$\tilde Z^{(1)}(q_{N_r};l_{1r},l_{2r})$ factor turns out to be
\begin{equation}
\tilde Z^{(1)}(q_{N_r};l_{1r},l_{2r})=
\frac{k_r^{\sigma_rl_{1r}}H_{(1)}(q_{N_r})[1+
(-1)^{l_{2r}}k_r]^{2l_{1r}}}
{k_r^{(3-2l_{1_r})/2}[1+(-1)^{l_{2r}}
k_r^{1/2}]^{4l_{1r}}}
Z^{(1)}(k_r;l_{1r},l_{2r})
\end{equation}
where the super-Schottky multipliers $k_s$
are defined in (19)
and  $Z^{(1)}(k;l_{1},l_{2})$ is
\begin{equation}
Z^{(1)}(k;l_{1},l_{2})=\prod_{p=1}^\infty
\frac{[1+(-1)^{2l_2}k^pk^{(2l_1-1)/2}]^8}{[1-k^p]^8}.
\end{equation}
In (109)  and (110) it is implied that
$l_j\in(0,1/2)$
where $j=1,2$. The $H_{(1)}(q_{N_r})$ factor
in (109) is
\begin{equation}
H_{(1)}(q_{N_r})=(u_r-v_r-\mu_r\nu_r)^{-1}.
\end{equation}

The terms in the square brackets in (105) originate
in the considered $Z_L^{(n)}(\{q_{N_s}\})$
the  multiplier $ Z_m^{(n)}(\{q_{N_s}\},L)$.
The above multiplier turns out to be
\begin{equation}
\ln Z_m^{(n)}(\{q_{N_s}\},L)=-5trace\ln(I-\hat
K^{(1)}+\hat\varphi\hat f)
\end{equation}
where  the $\hat K^{(1)}$ and $\hat\varphi\hat f$ operators
are the same as in (58).
For  superspin structures without the
odd genus-1 superspin ones eq.(112) is proved
in a rather
simple manner. In this case the discussed factor is
determined by the second term in (61).
As far as $\hat K_{ss}^{(1)}=0$,  the above term in
(61) for $r=s$ being considered, can be
rewritten as $\hat K^{(1)}\hat K\tilde K^{(1)}$.
So the above term
contributes to eqs.(100) as follows
\begin{equation}
-5\sum_{m_1\neq
r}\int\limits_{C_{m_1}}dt_1
\sum_{m_2}\int\limits_{C_{m_2}}\hat K(t_1,t_2)dt_2
\sum_{p=1,2}\int\limits_{C_p^{(r)}}K_r^{(1)}(t_2,t)dt
\partial_zK_r^{(1)}(t,t_1)Y_{N_r}(t)
\end{equation}
where the $C_m$ contours are defined as in (53).
The  integral over $t$ in (113) can
be compared with the following expression for $\partial_{N_r}
K_r^{(1)}(t_2,t_1)$:
\begin{equation}
\partial_{N_r}K_r^{(1)}(t_2,t_1)=
\sum_{p=1,2}\int\limits_{C_p^{(r)}}K_r^{(1)}(t_2,t)dt
E_{p,N_s}^{(0)}(K_r^{(1)}(t,t_1))
\end{equation}
where $E_{p,N_s}^{(0)}$ is given by (93) at $q=0$. To
obtain eq.(114) we represent  $\partial_{N_r}
K_r^{(1)}(t_2,t_1)$ in the form of the integral over $t$ along
the infinitesimal contour $C(z_2)$ that surrounds the $z_2$
point, the integrand being $K_r^{(1)}(t_2,t)\partial_{N_r}
K_r^{(1)}(t,t_1)$.  Using (92), we transform the above
integral to the integral (114) along $C_a^{(r)}$ and
$C_b^{(r)}$ contours, see also Appendix D.

We argue that the difference of (114) and the integral
over $t$ in (113) does not contributes in (113).
The simplest way to prove this
statement is to add in $<T_m>$ the
term $5D[<(DX)(DX)>]$. The above term is equal to zero because
$DX$ obeys the fermi statistics. Being substituted in
(89) and integrated by parts, the discussed term
transforms the integral over $t$ in (113) into the
integral (114).  So the integral over $t$ in (113)
can be replaced by $\partial_{N_r}K_r^{(1)}(t_2,t_1)$. So
(113) appears to be
$-5\partial_{N_r}trace\ln(I-\hat K^{(1)})$ that
proves eq.(112).

The  proof of eq.(112) for those
even genus-$n$ superspin structures where the odd genus-1
superspin structures present can be done in a like manner.
In this case the discussed
$ Z_m^{(n)}(\{q_{N_s}\},L)$ factor is
determined by all the  terms in (61)
except only the first term.To obtain the integral
representations of the derivatives with
respect to the moduli we use  (D20) and
(D21) of Appendix D. The contribution into
eqs.(100) due to the discussed terms turns out to be
\begin{equation}
5trace\left\{\left[I+\hat K-(I+\hat K)\varphi
V^{-1}\hat f-(I+\hat K)\varphi V^{-1}\hat f\right]
\partial_{N_r}(-\hat K^{(1)}+\hat\varphi\hat f)\right\}.
\end{equation}
On the other side, one can verify that
\begin{equation}
(I-\hat K^{(1)}+\hat\varphi\hat f)^{-1}=
I+\hat K-(I+\hat K)\hat
\varphi
V^{-1}\hat f-(I+\hat K)\hat\varphi V^{-1}\hat f.
\end{equation}
So eq.(115) can be rewritten as
$-5\partial_{N_r}trace\ln(I-\hat K^{(1)}+\hat\varphi
\hat f)$ that
proves eq.(112).

The calculation of $\tilde Z_{gh}^{(n)}(\{q_{N_s}\},L)$ from
eqs.(108) is performed by the kindred method.
We use (D22) and (D23) of Appendix D.
We use also (88) and (87) for
the calculation of the appropriate integrals. Moreover,
in this calculation
a number of terms disappears owing to the following identity
\begin{equation}
\sum_{N_r'}\sum_{p=1,2}\int\limits_{C_p^{(r)}}\chi_{N_r'}(t)dt
\left[E_{p,N_r}^{(-2)}(Y_{p,N_r'}(t))-
\partial_{q_{N_r}}Y_{p,N_r'}(t)+(-1)^{e(N_r')e(N_r)}
\partial_{q_{N_r'}}Y_{p,N_r}(t)\right]=0
\end{equation}
To prove (117) one uses that in (117), just as in
(96), the polynomials in the square brackets are not
higher degree in $(z,\theta)$, than degree-2.  So, to calculate
the left side of (117), we again can use (A3)
and (88).  The most
simple way to perform this calculation is to go from $t$ to the
$t_r$ variables (23). As the result, eq.(117) arises.
The result of the calculation of $\tilde
Z_{gh}^{(n)}(\{q_{N_s}\},L)$  is given as follows
\begin{equation}
\ln\tilde Z_{gh}^{(n)}(\{q_{N_s}\},L) =
\ln sdet[M(\{\sigma_p\})]+trace\ln(I-\hat S_\sigma^{(1)})
\end{equation}
where the $M(\{\sigma_p\})$ matrix is defined in
(77), the $\hat S_\sigma^{(1)}$ operator is the same
as in (84)  and the superdeterminant
$sdet\tilde U$ of any  $\tilde U$ matrix is  defined as
\begin{equation}
sdet\tilde U=\frac{\det
\tilde U_{(bb)}}{\det \tilde U_{(ff)}}\det
[I-\tilde U_{(bb)}^{-1}
\tilde U_{(bf)}U_{(ff)}^{-1}\tilde U_{(fb)}]
\end{equation}
where $\tilde U_{(bb)}$, $\tilde U_{(bf)}$,
$\tilde U_{(fb)}$ and $\tilde U_{(ff)}$ are submatrices
forming the above $\tilde U$ matrix.  The index $b$ labels
boson components and the index $f$ labels the fermion ones.
The $M_{N_sN_r'}(\{\sigma_p\})$
element of the  $M(\sigma_p)$ matrix (77)
are found to have the following form
\begin{equation}
M_{N_sN_r'}(\{\sigma_p\})=\delta_{rs}\tilde I_{N_s}
M_{(1)N_sN_r'}(\sigma_s)+
M_{N_sN_r}'(\{\sigma_p\})
\end{equation}
where $M_{N_sN_r}'(\{\sigma_p\})\rightarrow0$, if
the handle labeled by $s$ ( or by $r$ ) goes to infinity
provided that the above handle is associated
with the even genus-1 spin structure.
Furthermore, the $\tilde I_{N_s}$ is defined in
(120) as
\begin{eqnarray}
\tilde
I_{\mu_s}=1+2l_{1s}-2l_{1s}l_{2s}[1-\sigma_s],\quad \tilde
I_{\nu_s}=1+2l_{1s}-2l_{1s}l_{2s}[1+\sigma_s]\nonumber \\
{\rm and}\quad \tilde I_{N_s}=1 \quad{\rm for}
\quad N_s=k_s,u_s,v_s.
\end{eqnarray}
The $M_{(1)N_sN_r'}(\sigma_s)$ matrix in (120)
has the following form
\begin{equation}
M_{(1)N_sN_r'}(\sigma_s)=
\delta_{rs}M_{N_sN_s'}^{(s)}(\sigma_s)
\end{equation}
The  $M^{(s)}(\sigma_s)$ matrix in (122)
is formed by the
$M_{(ff)}^{(s)}(\sigma_s),
M_{(bf)}^{(s)}(\sigma_s),M_{(fb)}^{(s)}(\sigma_s)$ and
$M_{(ff)}^{(s)}(\sigma_s)$ submatrices.
The index $b$ labels
boson components and the index $f$ labels the fermion ones.
The above submatrices are defined as
follows. Firstly, $M_{(bb)}^{(s)}(\sigma_s)=I$ and
$M_{(fb)}^{(s)}(\sigma_s)=0$. Besides, the elements of the
$M_{(bf)}^{(s)}(\sigma_s)$ matrix are given by
\begin{eqnarray}
M_{b\mu_s}^{(s)}(\sigma_s)=
\frac{(2+\sigma)M_{b\mu_s}^{(s)}(\sigma_s)}{2-\sigma}=-
\frac{l_{1s}(\mu_s-\nu_s)(1+2\sigma_s)}{2}\quad
{\rm for}\quad b=u_s,v_s\nonumber \\
{\rm and}\quad M_{k_sf}^{(s)}(\sigma_s)=0.
\end{eqnarray}
The elements $M_{ff'}(k_s,\sigma_s)$ of the above
$M_{(ff)}^{(s)}(\sigma_s)$ matrix are defined as follows
\begin{eqnarray}
M_{\mu\mu}(k,-1)=M_{\mu\nu}(k,-1)=M_{\nu\nu}(k^{-1},1)=
M_{\nu\mu}(k^{-1},1)=M_\mu(k) \quad  {\rm and}\nonumber \\
M_{\nu\nu}(k,-1)=-3M_{\nu\mu}(k,-1)=M_{\mu\mu}(k^{-1},1)=
-3M_{\mu\nu}(k^{-1},1)=M_\nu(k)
\end{eqnarray}
where $M_\mu(k)$ and $M_\nu(k)$ are given by
\begin{equation}
M_\mu(k)=
1-2l_{1s}+\frac{(-1)^{2l_{2s}}l_{1s}\sqrt k}{[1+
(-1)^{2l_{2s}}\sqrt k]}\quad
{\rm and}\quad M_\nu(k)=
1-2l_{1s}+\frac{3l_{1s}[1+(-1)^{2l_{2s}}k])}
{[1+ (-1)^{2l_{2s}}\sqrt k]}.
\end{equation}
In (124) and (125) the index $s$
is omitted. In (121)-(125) it is implied that
$l_{1s}\in(0,1/2)$ and $l_{2s}\in(0,1/2)$.
In the following consideration
it will be convenient to
rewrite (118) as
\begin{equation}
\tilde Z_{gh}^{(n)}(\{q_{N_s}\},L) =
sdet[M(\{\sigma_p\})U^{-1}(\{\sigma_p\})]
Z_{gh}^{(n)}(\{q_{N_s}\},L)
\end{equation}
where  $Z_{gh}^{(n)}(\{q_{N_s}\},L)$ is defined by
\begin{equation}
\ln
Z_{gh}^{(n)}(\{q_{N_s}\},L) =
trace\ln(I-\hat S_\sigma^{(1)})+
\ln sdet[U(\{\sigma_p\})].
\end{equation}
The
$U(\{\sigma_p\})$ matrix in (126) and
in (127) is defined as
\begin{equation}
U(\{\sigma_p\})=M(\{\sigma_p\})M_{(1)}^{-1}(\{\sigma_p\}).
\end{equation}
In (128) the $M_{(1)}(\{\sigma_p\})$
matrix is defined by (122).
One can verify that the
$U_{N_s,N_r}(\{\sigma_p\})$ elements of the
$U(\{\sigma_p\})$ matrix  (128)
have the following form
\begin{equation}
U_{N_s,N_r}(\{\sigma_p\})=\tilde I_{N_s}\delta_{N_sN_r}+
\hat U_{N_s,N_r}(\{\sigma_p\})
\end{equation}
where $\hat U_{N_s,N_r}(\{\sigma_p\})$ decrease,
if at least one of two handles labeled by $s$
and by $r$ goes away to infinity
provided that the above handle is associated
with the even genus-1 spin structure..
And  $\tilde I_{N_s}$ is given by (121).

Collecting together all the obtained factors in
$Z_L^{(n)}(\{q_{N_s}\})$, we write down the desired
$Z_L^{(n)}(\{q_{N_s}\})$ factor in (90) as
\begin{equation}
Z_L^{(n)}(\{q_{N_s}\})=
\tilde Z^{(n)}(\{q_{N_s}\},L)
H(\{q_{N_s}\}) \prod_{s=1}^n
\frac{(-1)^{2l_{1s}+2l_{2s}-1}16^{2l_{1s}}
Z^{(1)}(k_s;l_{1s},l_{2s})}{k_s^{(3-2l_{1s})/2}}
\end{equation}
where $Z^{(1)}(k_s;l_{1s},l_{2s})$ is given by
(110) and $\tilde Z^{(n)}(\{q_{N_s}\},L)$ is defined by
\begin{equation}
\tilde Z^{(n)}(\{q_{N_s}\},L)=
Z_m^{(n)}(\{q_{N_s}\},L)
Z_{gh}^{(n)}(\{q_{N_s}\},L)
\end{equation}
where $Z_m^{(n)}(\{q_{N_s}\},L)$ is defined by
(112) and $Z_{gh}^{(n)}(\{q_{N_s}\},L)$ is given by
(127).  Furthermore, the $H(\{q_{N_s}\})$ factor in
(130) is defined as
\begin{equation}
H(\{q_{N_s}\})=(u_1-u_2)(v_1-u_2)
\left[1-\frac{\mu_1\mu_2}{2(u_1-u_2)}-
\frac{\nu_1\mu_2}{2(v_1-u_2)}\right]
\prod_{s=1}^n(u_s-v_s-\mu_s\nu_s)^{-1}
\end{equation}
In (132) it is assumed that
$u_1,v_1,u_2,\mu_1$ and $\nu_1$  are fixed to be
the same for all the genus-$n$ supermanifolds and,
therefore, they are not the moduli.
Apart of the normalization  factor $(u_1-u_2)(v_1-u_2)$,
eq. (132) presents product $H_o$ and $H_1$
factors defined by (104) and (111).
As it has been noted already,
the above factor is calculated [11,12] from the
condition that the superstring
amplitudes are independent of a choice of $\{q_{N_o}\}$ set.
The  $u_1,v_1,u_2,\mu_1$ and $\nu_1$ fixed
parameters can be changed by fractionally linear
supersymmetrical transformations
(18). In (130) only the $H$ factor
(132) is changed
under these transformations. The condition that
the change of $H$ is
compensated by the changes of the differentials
in (15) just gives the above normalization factor
in (132) . In the calculation of
the discussed normalization factor  one is due to
take into account that the above transformations
depend not only on $u_1,v_1,u_2,\mu_1$
and $\nu_1$, but also on $\mu_2$, which is the
variable of the integration in (15).

In (130) the normalization factors
$(-1)^{2l_{1s}+2l_{2s}-1}16^{2l_{1s}}$ are taken into account.
To derive these factors in the case when
$l_{1s}l_{2s}=0$ we move away the handle labeled
by the $s$ index. In this case the genus-$n$ amplitude
is due to be proportional to the
one loop amplitude [17].  One can verify that in the
$l_{1s}l_{2s}=0$ case the dependence on $q_{N_s}$
Schottky parameters
disappears in both (112) and (118).
The comparison (130) with  [17] gives the above
normalization factors in (130).

Eq.(130) implies also the $16^2$ factor
for every pair of the odd genus-1 spin  structures
presenting in the given genus-$n$ superspin structure.
To derive this factor one can note that the superspin
structure containing a pair of
the odd genus-1 spin structures labeled, say, by
the $r$ and $s$ indices, can be derived by the
supermodular transformation of the
superspin structure containing a pair of the handles with
$l_{1r}=l_{12}=1/2$ and $l_{2r}=l_{2s}=0$. In the
case of zero odd Schottky parameters the above
transformation implies only adding
$\pm2\pi$ to the phase of   $u_r-u_s,v_r-v_s,u_r-v_s$ or of
$v_r-u_s$.  In the case of arbitrary odd parameters
that must be considered to derive the
desired factors in (130), the discussed
transformation includes, in addition, a change of the Schottky
parameters. So the Jacobian in (15) arises under
the discussed transformation. Fortunately, to derive
the  normalization factors in (130) one can consider
the case when both $k_1$ and $k_2$ tend to zero.
In this limit the discussed supermodular transformation does
not change of the Schottky parameters and, therefore in
the considered limit the
Jacobian of this  transformation is equal to
unity. In this case the requirement of the invariance
of (15) under the considered supermodular
transformation gives the above $16^2$ factor
announced in (130).

Though $Z_{gh}^{(n)}(\{q_{N_s}\},L)$ in (131)
being given by
(127), is formed by the multipliers
depending on a choice of the $\{\sigma_p\}$ set, the
above $Z_L^{(n)}(\{q_{N_s}\})$ factor does not depend on
$\{\sigma_p\}$. Indeed, as
it has been shown in Sec.V, the ghost vacuum correlator
(62) is determined  uniquely  and, therefore, this
correlator is independent of a choice of the  $\{\sigma\}$ set.
So, $Z_L^{(n)}(\{q_{N_s}\})$  is independent of a choice of
$\{\sigma\}$. Therefore, one concludes that
$Z_{gh}^{(n)}(\{q_{N_s}\},L)$ is also independent of
$\{\sigma_p\}$.

Eq.(130) gives the desired $Z_L^{(n)}(\{q_{N_s}\})$
holomorphic multiplier in the partition function (90).
To obtain the final result, in the next Section we  yet
calculate  both
$Z_{gh}^{(n)}(\{q_{N_s}\},L)$ and $Z_m^{(n)}(\{q_{N_s}\},L)$
multipliers in (131) in the terms of  moduli and of
Green functions.

\section{ Final expressions for the superstring amplitudes}

To obtain the explicit formula for the
$\tilde Z^{(n)}(\{q_{N_s}\},L)$ factor in (130) we
rewrite this factor  as
\begin{equation}
\tilde Z^{(n)}(\{q_{N_s}\},L)=
\tilde Z_{0}^{(n)}(\{q_{N_s}\},L)
\Upsilon_m^{(n)}(\{q_{N_s}\},L)
\Upsilon_{gh}^{(n)}(\{q_{N_s}\},L)
\end{equation}
where $\tilde Z_{0}^{(n)}(\{q_{N_s}\},L)$ is the value of the
considered factor at zero odd Schottky parameters. Moreover,
$\ln\Upsilon_m^{(n)}(\{q_{N_s}\},L)$ and
$\ln\Upsilon_{gh}^{(n)}(\{q_{N_s}\},L)$  present the terms
proportional to the odd Schottky parameters in (112)
and in (127), respectively. So
$\ln\Upsilon_m^{(n)}(\{q_{N_s}\},L)$ can be given as
\begin{equation}
\ln\Upsilon_m^{(n)}(\{q_{N_s}\},L)=\ln\left[1-\delta\hat
K^{(1)}+\delta[\hat\varphi\hat f]+\Delta_m\right]
\end{equation}
where $\delta\hat K^{(1)}$ and $\delta[\hat\varphi\hat f]$ denotes
those terms in $\hat K^{(1)}$ and $\hat\varphi\hat f$, respectively,
which are proportional to the odd Schottky parameters. Furthermore,
the  $\Delta_m$ integral operator is formed by the
$\{\Delta_m^{(p)}\}$ set of the $\Delta_m^{(p)}$ integral operators,
the  "kernels" being $\Delta_m^{(p)}(t,t')dt'$.
As in (59), we  define the kernel together
with the differential $dt'$. Every the $\Delta_m^{(p)}$ integral
operator being applied to a function of $t'$, performs integrating
over $t'$ along the $C_p$ contour. The  $C_p$ contours are
the same as in  (56). Moreover, one can verify that
\begin{eqnarray}
\Delta_m^{(p)}(t,t')=\int\limits_{C_p}\left[
K_{(o)}(t,t_1)+
\int\limits_{C_p}2K_{(o)}(t,t_2)dt_2\varphi_{(o)p}(t_2)
f_{(o)p}(t_1)\right]dt_1\delta [\varphi_p(t_1)f_p(t')]-\nonumber \\
\sum_{r\neq
p}\int\limits_{C_r}\left[ K_{(o)}(t,t_1)+
\int\limits_{C_r}2K_{(o)}(t,t_2)dt_2\varphi_{(o)r}(t_2)
f_{(o)r}(t_1))\right]dt_1\delta K_r^{(1)}(t_1,t')
\end{eqnarray}
where  $K_{(o)}(t,t')$, $\varphi_{(o)r}$ and
$f_{(o)r}$ denote, respectively, $K(t,t')$, $\varphi_r$ and $f_r$
calculated at all  odd Schottky parameters to be equal to zero.
Eq.(135) follows from (59) and from
(116).

Furthermore, $\ln\Upsilon_{gh}^{(n)}(\{q_{N_s}\},L)$
in (133) can be given in a
quite like manner in the terms of both the $S_{(o)\sigma}(t,t')$
Green function (78) and the proportional to odd
parameter terms in $\hat S_\sigma^{(1)}$.
Besides, in $\ln\Upsilon_{gh}^{(n)}(\{q_{N_s}\},L)$  the $\ln
sdet[U(\{\sigma_p\})U_{(o)}^{-1}(\{\sigma_p\})]$ term
presents, as it follows from (127). The
$U_{(o)}(\{\sigma_p\})$ is defined to be $U(\{\sigma_p\})$
at zero odd moduli. The fermion part $S_{(f)\sigma}(t,t')$ of
$S_{(o)\sigma}(t,t')$ is  calculated in the terms of the
$G_{(\sigma)} (z,z')$
Green function (69) by (79). So
$\ln\Upsilon_{gh}^{(n)}(\{q_{N_s}\},L)$ turns
out to be given in
the terms of the following
$G^0(t,t';\{\sigma_p\})$ Green function
\begin{equation}
G^0(t,t';\{\sigma_p\})=
G_b(z,z')\theta'+\theta G_{(\sigma)} (z,z')
\end{equation}
where $G_b(z,z')$
is defined by (68) and  $G_{(\sigma)} (z,z')$  is defined by
(69). The final result for
$\ln\Upsilon_{gh}^{(n)}(\{q_{N_s}\},L)$ is found to be
\begin{eqnarray}
\ln\Upsilon_{gh}^{(n)}(\{q_{N_s}\},L)=
trace\ln\left[I-\delta\hat S_\sigma^{(1)}+
\Delta_{gh}(\{\sigma_p\})\right]+ \nonumber \\
\ln
sdet[U(\{\sigma_p\})U_o^{-1}(\{\sigma_p\}) U'(\{\sigma_p\})]
\end{eqnarray}
where $\delta\hat S_\sigma^{(1)}$ is referred to those terms in
$\hat S_\sigma^{(1)}$, which are proportional to the odd
Schottky parameters. Moreover,$U_{(o)}(\{\sigma_p\})$ is
$U(\{\sigma_p\})$ at zero odd moduli. Furthermore, the
$\Delta_{gh}(\{\sigma_r\})\}$ integral operator is formed by the
$\{\Delta_{gh}^{(p)}(\{\sigma_r\})\}$ set of the
$\Delta_{gh}^{(p)}(\{\sigma_r\})$ integral operators,
the  kernels being $\Delta_{gh}^{(p)}(\{\sigma_r\})(t,t')dt'$.
We again define the kernel together
with the differential $dt'$. Every the
$\Delta_{gh}^{(p)}(\{\sigma_r\})$ integral
operator being applied to a function of $t'$, performs integrating
over $t'$ along the $C_p$ contour that is
the same as in  (56). Furthermore,
the $U_{N_rN_s}'(\{\sigma_p\})$ elements
of the $U'(\{\sigma_p\})$ matrix are defined as
\begin{eqnarray}
U_{N_rN_s}'(\{\sigma_p\})=I+\sum_{p\neq
q}\int\limits_{C_p}\Psi_{\sigma,N_r}^{(0)}(t)dt
\int\limits_{C_q}\delta
S_{\sigma,p}^{(1)}(t,t_1)dt_1 \times \nonumber \\
\sum_h
\int\limits_{C_h}\tilde\Delta^{(h)}(t_1,t_2)
dt_2\int\limits_{C_s}G^0(t_2,t';\{\sigma_p\})dt'
\hat Y_{N_s}^{(1)}(t')
\end{eqnarray}
where $\Psi_{\sigma,N_r}^{(0)}(t)$ are $3/2$-supertensors
defined by (71). And
$\tilde\Delta^{(h)}(t_1,t_2)dt_2$ present the kernels of
the $\tilde\Delta^{(h)}$ integral operators.
The $\{\tilde\Delta^{(h)}\}$ set of these operators forms
the $\tilde\Delta$  operator that can be given as
\begin{equation}
\tilde\Delta=[I+\Delta_{gh}(\{\sigma_p\})]^{-1}
\end{equation}
where the $\Delta_{gh}(\{\sigma_p\})$ operator
is the same as in (137). The  kernels
$\Delta_{gh}^{(p)}(\{\sigma_r\})(t,t')dt'$
of the above $\Delta_{gh}(\{\sigma_p\})$ operator are defined by
\begin{eqnarray}
\Delta_{gh}^{(p)}(\{\sigma_r\})(t,t')=-\sum_{r\neq
p}\int\limits_{C_r}G^0(t,t_1;\{\sigma_q\})dt_1\delta
S_\sigma^{(1)}(t_1,t').
\end{eqnarray}
Eqs. (134)-(140) allow to calculate both
$\ln\Upsilon_m^{(n)}(\{q_{N_s}\},L)$
and $\ln\Upsilon_{gh}^{(n)}(\{q_{N_s}\},L)$, at
least in the form of the series over
the odd Schottky parameters.

So, to obtain the explicit formulae for the holomorphic
factors  in the partition functions (90)
one is due  to calculate at zero odd Schottky
parameters the $\tilde
Z^{(n)}(\{q_{N_s}\},L)$ factor in (130).
This factor is referred as $\tilde
Z_0^{(n)}(\{q_{N_s}\},L)$ in (133).
We show now that in the case of zero all the odd Schottky
parameters the discussed  holomorphic
factors in (90) can  be calculated
explicitly.  The above holomorphic
factors at zero odd Schottky parameters being known, one  can
derive the desired $\tilde Z_0^{(n)}(\{q_{N_s}\},L)$ factor
in (133). So the resulting holomorphic factors in
(90) can be given as follows
\begin{eqnarray}
Z_L^{(n)}(\{q_{N_s}\})= Z_{0(m)}^{(n)}(\{k_s,u_s,v_s\},L)
Z_{0(gh)}^{(n)}(\{k_s,u_s,v_s\},L)\times \nonumber \\
H(\{q_{N_s}\})
\Upsilon_m^{(n)}(\{q_{N_s}\},L)
\Upsilon_{gh}^{(n)}(\{q_{N_s}\},L)
\end{eqnarray}
where $H(\{q_{N_s}\})$ is given by (132),
$\Upsilon_m^{(n)}(\{q_{N_s}\},L)$ is defined by
(134) and $\Upsilon_m^{(n)}(\{q_{N_s}\},L)$
is defined by (137).  The
$Z_{0(m)}^{(n)}(\{k_s,u_s,v_s\},L)$ factor in (141)
is due to the string fields and
$Z_{0(gh)}^{(n)}(\{k_s,u_s,v_s\},L)$ is due to the ghost fields.
Both these factors being calculated at zero  odd Schottky
parameters, depend only on the even Schottky ones.
To calculate  the discussed factors, we use eqs.(94)
for $N_r=(k_r,u_r,v_r)$. In the integrand in
(94), there are the terms proportional
to the derivatives with respect to odd modui of
the $Y_{N_r}$ polynomials. We calculate the above
derivatives using eqs.(A1) of Appendix A.
After the discussed
derivatives to be calculated, we  take all the odd Schottky
parameters in (94) to be equal to zero. In this case
the $I_{(an)}$ term disappears. Moreover, the integrals over the
$C_a^{(r)}$ paths vanish because $Y_{a,u_r}=Y_{a,v_r}=
Y_{a,k_r}=0$ at $\mu_r=\nu_r=0$.  The integral of
$W_{p,N_r}^{(o)}(t)$  disappears, too. It is useful to note that
at zero odd Schottky parameters  the boson field contributions
and the fermion field contributions can calculate separately
from each other. For the boson field contributions one can use
the boson string calculations [10,13,26], but
the fermion field contributions need in a special consideration.

In the calculation of $Z_{0(gh)}^{(n)}(\{k_s,u_s,v_s\},L)$
we express the $G(t,t')$ ghost Green function (63) in
the terms of $\hat G^0(t,t';\{\sigma\})$ where the
$\hat G^0(t,t';\{\sigma\})$ Green functions is defined as
\begin{equation}
\hat G^0(t,t';\{\sigma\})= \frac{1}{2}[G^0(t,t';\{\sigma_p\})+
G^0(t,t';\{-\sigma_p\})]
\end{equation}
where $ G^0(t,t';\{\sigma_p\})$ are defined by (136).
To employ $\hat G^0(t,t';\{\sigma\})$, it is much more
convenient, than to employ $G^0(t,t';\{\sigma_p\})$ because
in this case the terms proportional to $\sigma_p$ disappear
in the calculation of $Z_{0(gh)}^{(n)}(\{k_s,u_s,v_s\},L)$.
To express $G(t,t')$ in the terms
of $\hat G^0(t,t';\{\sigma\})$,
we use (67), (73) and (136), as well.
The integrals over $z''$ in (73) give
raise the integrals in (94) over both $t$ and $t''$.
In these integrals the integration over $t''$ is performed
before the integration over $t$, but we change
the above order of the integration and perform the integration
over $t$ before the integration over $t''$, as we made it  in
Sec.VI above. In the terms associated with $s=r$ in the sum over
$s$ in (73) there is implied that the pole at$z''=z$ is
situated outside of the $C_b{(r)}$ contour on the $z''$ complex
plane. But we  accommodate the above pole inside the $C_b{(r)}$
contour adding simultaneously  the suitable
terms to cancel the contribution of the discussed pole.

In the  right side of the discussed equations
one observes a large number of  terms, which  can be rewritten
in the form of the integral $I'$ along the $C_q$ contours as
\begin{equation}
I'=\frac{1}{2}\sum_{q=1}^n\int\limits_{C_q}dz
[G_{(\sigma)}(z,z)-G_{(-\sigma)}(z,z)]
\sum_{s}2l_{1s}\sigma_s\partial_{q_{N_r}}J_{(o)s}(z)
\end{equation}
where the $C_q$ contours are defined in (56) and, as it
is usual, $G_{(\sigma)}(z,z)$ is defined to be the limit of
$G_{(\sigma)}(z,z')$ at $z\rightarrow z'$, the singular term
$(z-z')^{-1}$ being omitted. The $G_{(\sigma)}(z,z')$ Green
function is given by (73). Furthermore,
$G_{(-\sigma)}(z,z)$ is obtained from $G_{(\sigma)}(z,z)$
by the $\sigma_p\rightarrow-\sigma_p$ replacement
for every $\sigma_p$. The integrand in (143) has no
singularities outside the $C_q$ contours and, in addition,
it tends to zero more rapidly than $z^{-1}$ at
$z\rightarrow\infty$. Therefore, $I'=0$.
The rest of the terms on the right side of every of the
considered equations can be written down in the form of
the derivative with respect to $q_{N_r}$ of a function of the
moduli, which turns out to be the same for all the equations
discussed. To verify this statement one needs the derivatives
with respect to moduli of the multipliers assigned to  group
products of the basic Schottky group elements. The above
derivatives can be obtained from the formulae given in Appendix
E. The discussed expressions for the desired derivations have
been already used in [10,13,26] though
in [10,13,26] the above expressions were not given in
an explicit form.

In the calculation of
the $Z_{0(m)}^{(n)}(\{k_s,u_s,v_s\},L)$ factor
in (141) we use the
$R_{(o)}$  Green function (44). Furthermore, we employ
the $Z_{0(m)}^{(n)}(\{k_s,u_s,v_s\},L_0)$
factor [11,12,22]
assigned to the $L_0=\bigcup_s(l_{1s}=0 ,l_{2s}=1/2)$
superspin structure  to express those contributions to every
$Z_{0(m)}^{(n)}(\{k_s,u_s,v_s\},L)$, which do not
depend on the superspin structure considered.
In this case the desired
$Z_{0(m)}^{(n)}(\{k_s,u_s,v_s\},L)$ factors
can be written down as
\begin{equation}
Z_{0(m)}^{(n)}(\{k_s,u_s,v_s\},L)=
\frac{\Theta^5[l_1,l_2](0|\omega^{(o)})}
{\Theta^5[\{0\},\{1/2\}] (0|\omega^{(o)})}\prod_{(k)}
\prod_{m=1}^\infty \frac{(1-k^{m-1/2})^{10}}{(1-k^m)^{10}}
\end{equation}
where $\Theta$ is
the theta function. The $\Theta$ in the denominator associates with
the $S_0$ spin structure. The period matrix $\omega^{(o)}$ being
calculated at zero odd Schottky parameters, is given by
eq.(B9) of Appendix B.  The product over $(k)$ is taken over
all the multipliers of the Schottky group (17), which are
not powers of other the ones. To obtain (144) we employ
eq.(D18) of Appendix D for the derivatives with respect to the
moduli of the period matrix. Besides, we use the equations of
Appendix E for the derivatives with respect to moduli of the
multipliers assigned to Schottky group products.
In the calculation of $Z_{0(gh)}^{(n)}(\{k_s,u_s,v_s\},L)$
we employ also relations (87) and identity (117), as
well. The desired $Z_{0(gh)}^{(n)}(\{k_s,u_s,v_s\},L)$ factor in
(141) turns out to be
\begin{eqnarray}
Z_{0(gh)}^{(n)}(\{k_s,u_s,v_s\},L)=\frac{\exp[-\pi i
\sum_{j,r}l_{1j}l_{1r}\omega_{jr}^{(o)}]}
{\sqrt{\det M_{(o)}(\{\sigma_p\})
\det M_{(o)}(\{-\sigma_p\})}}
\left(\prod_{s=1}^nZ_0(k_s;l_{1s},l_{2s})\right)
\times\nonumber \\
\prod_{(k)}\prod_{m=1}^\infty\frac{(1-k^{m+1})^2}
{[1-\Lambda(k,\{\sigma_p\})k^{m+1/2}]
[1-\Lambda(k,\{-\sigma_p\})k^{m+1/2}]}
\end{eqnarray}
where, as in (144), the product over $(k)$ is taken
over all the multipliers of the Schottky group (17),
which are not powers of the other ones and
\begin{equation}
\Lambda(k,\{\sigma_p\})=\exp\Omega_{\Gamma_{(k)}}(\{\sigma_p\})
\end{equation}
where $\Omega_{\Gamma_{(k)}}(\{\sigma_p\})$
is given by (70) for those
group products of the basic Schottky transformations, which
have the multiplier to be equal $k$. The
$Z_0(k_s;l_{1s},l_{2s})$ factors in (145) are defined by
\begin{equation}
Z_0(k_s;l_{1s},l_{2s})=
\frac{(-1)^{2l_{1s}+2l_{2s}-1}(1-k_s)^2}{4^{2l_{1s}}k_s^{3/2}
[1+(-1)^{2l_{2s}}\sqrt k_sk_s^{l_{1s}}]^{2-2l_{1s}}}.
\end{equation}
Furthermore, the $M_{(o)}(\{-\sigma_p\})$ matrix in (145)
is given by
\begin{equation}
M_{(o)}(\{\sigma_p\})=\tilde M(\{\sigma_p\})
M_{(1,0)}^{-1}(\{\sigma_p\})
\end{equation}
where the $\tilde M(\{\sigma_q\})$ matrix is defined by
(74) and $M_{(1,0)}(\{\sigma_p\})$ is
$M_{(1)}(\{\sigma_p\})$ taken at zero odd Schottky parameters.
The above $M_{(1)}(\{\sigma_p\})$ matrix is the same as in
(128).  It is useful to remind
that in both (144) and (144)  the $k$ multipliers
are calculated at zero odd Schottky parameters. The
$(-1)^{2l_{1s}+2l_{2s}-1}2^{-2l_{1s}}$ normalization
factors in (147) is determined to obtain the
$(-1)^{2l_{1s}+2l_{2s}-1}16^{2l_{1s}}$
normalization factors in (130).

Eqs.(141) being added by (134), (135) and
(137)-(148),
solve the problem of the calculation of
the $Z_L^{(n)}(\{q_{N_s}\})$ holomorphic multipliers
in the partition functions. The another form of the above
holomorphic multipliers is presented by eqs. (130)
together with (112), (127), (132)
and (131).
The discussed $Z_L^{(n)}(\{q_{N_s}\})$ holomorphic
multipliers to be known, the partition functions are
determined by (90). In this
case the multi-loop amplitudes are calculated by
(15). To calculate  the vacuum expectations of
the vertex products in  (15) we use the
$\hat X_{L,L'}(t,\overline t;t',\overline t')$
vacuum correlator (31) of the scalar superfields.
The above correlators are calculated in the terms of the
holomorphic Green functions $R_L(t,t')$,  the $J_r(t;L)$
functions and the period
matrix $\omega(\{q_N\};L)$. At zero odd Schottky
parameters, the above functions are given by (B8)
and the period matrix is
given by (B9).  The depending on odd Schottky
parameter terms in $J_r(t;L)$ and $\omega(\{q_N\};L)$
are taken into account by
(54) and by (55).  Furthermore, at zero odd
Schottky parameters, the holomorphic Green function of the
string fields is given by (44) together with (B7)
and (45).  The depending on odd Schottky
parameter terms in  the discussed holomorphic
functions can be calculated by means of (53).
In the case when all the $l_{1s}$ characteristics
are equal to zero the partition functions, as well as the Green
functions can be given in the much simpler
form [11,12,22].
The integration region in (15) is determined
by  the requirement of the supermodular invariance.

The investigation of the obtained multi-loop amplitudes
is planned to perform in an another place. In this paper we
only touch shortly the divergency problem.

Owing to the supermodular invariance  one can
exclude from this region of the integration in (15) those
domains where some of the Schottky group multipliers $k$ are
near to unity: $k\approx 1$. Indeed, modulo of supermodular
transformations, these  domains are equivalent to those  where some
of $k_j$ are small: $k_j\approx 0$.
At $k_j\rightarrow 0$
we see from eq.(132) and (141) that
$Z_L^{(n)}(\{q_{N_s}\})\sim k_j^{-1}$
for $l_{1j}=1/2$ and
$Z_L^{(n)}(\{q_{N_s}\})\sim k_j^{-3/2}$
for  $l_{1j}=0$.
However, in the sum (15) over $L$ the above singularity
$k_j^{-3/2}$ is reduced to $k_j^{-1}$. Besides, we have
the factor $(\ln|k|)^{-5}$ due to the non-holomorphic factor in
the partition functions (90).
As the result, the integral over $k_j$ in(15) appears to
be finite at $k_j\rightarrow 0$.

Nevertheless, the problem of the finiteness of the considered
theory needs a further study. It follows from (132)
that, beside the above singularities at $k_j\rightarrow 0$, every
$Z_L$ has also the singularities at
$u_j-v_j\rightarrow 0$. One can interpret the above limit as
the moving of the $j$-handle away from the others.
The contribution to $A_n$ from the region where
$u_j-v_j\rightarrow 0$ appears to be proportional to
\begin{equation}
\int\frac{d(Reu_j)d(Rev_j) d(Imu_j)d(Imv_j) d\mu_jd\nu_j
d\overline\mu_jd\overline\nu_j}{|u_j-
v_j-\mu_j\nu_j|^2}Z_{n-1}A_1
\end{equation}
where $Z_{n-1}$ is
the genus-(n-1) vacuum amplitude and $A1$
is the genus-1 amplitude. One can see that the
integral (149) has   uncertainty, if
$Z_{n}\neq 0$ for all $n>1$.
The uncertainties of the same type
arise also from the other regions, which correspond
to the moving of the handles away from each other.
The equality $Z_n=0$  is expected [9], if the
discussed theory possesses the space-time
supersymmetry, as well as the world-sheet one, but till
now we do not know an explicit proof of the statement
that the Ramond-Neveu-Schwarz
superstring really possesses the space-time supersymmetry.  This
problem, as well as the divergency problem in the
Ramond-Neveu-Schwarz superstring theory is planned
to be consider in an another paper.
\begin{center}
{\bf Acknowledgments}
\end{center}
The research described in this publication was made possible
in part by Grant No. NO8000 from the International
Science Foundation and in part by Grant No. 93-02-3147 from
the Russian Fundamental Research
Foundation.

\appendix
\section{Degree-2 polynomials in $(z,\theta)$}
The $Y_{b,k}$,  $Y_{b,u}$ and $Y_{b,\mu}$  polynomials can be
written down as
\begin{eqnarray}
Y_{b,k_s}(t)=\frac{Y_{b,k_s}^{(0)}(t_s)}
{Q_{\tilde\Gamma_s}^2(t_s)},
\quad Y_{b,u_s}(t)=
\frac{Y_{b,u_s}^{(0)}(t_s)(u_s-v_s)}
{Q_{\tilde\Gamma_s}^2(t_s)
(u_s-v_s-\mu_s\nu_s)}+
\frac{(\mu_s-\nu_s)Y_{b,\mu_s}^{(0)}(t_s)}
{Q_{\tilde\Gamma_s}^2(t_s)(u_s-v_s)} \nonumber\\
{\rm and}\quad\quad
Y_{b,\mu_s}(t)=\frac{(u_s-v_s-\mu_s\nu_s/2))
Y_{b,\mu_s}^{(0)}(t_s)}
{Q_{\tilde\Gamma_s}^2(t_s)(u_s-v_s)} +
\frac{\mu_sY_{b,u_s}^{(0)}(t_s)}
{Q_{\tilde\Gamma_s}^2(t_s)}
\end{eqnarray}
where both the $\tilde\Gamma_s$ mapping and $t_s=t_s(t)$
are defined in Sec.III by eq.(23). Furthermore,
\begin{eqnarray}
Y_{b,k_s}^{(0)}(t)=\frac{(z-u_s)(z-v_s)}{k_s(u_s-v_s)},\quad
Y_{b,u_s}^{(0)}(t)=
\frac{(1-k_s)(z-v_s)^2}{k_s(u_s-v_s)^2} \nonumber \\
{\rm and}\quad
Y_{b,\mu_s}^{(0)}(t)=\frac{2\theta(1-\sqrt k_s)(z-v_s)}
{\sqrt k_s(u_s-v_s)}.
\end{eqnarray}
One can see from (A1) that in the  $\mu_s=\nu_s=0$ case
the discussed polynomials are reduced to $Y_{b,k_s}^{(0)}$,
$Y_{b,u_s}^{(0)}$ and $Y_{b,\mu_s}^{(0)}$, respectively.
The $Y_{b,v_s}$ and $Y_{b,\nu_s}$ polynomials can be obtained
from $Y_{b,u_s}$ and $Y_{b,\mu_s}$, respectively, by the
replacements $k\rightarrow 1/k,u\rightarrow v, v\rightarrow
u,\mu\rightarrow\nu$ and $\nu\rightarrow\mu$.

Every degree-2  polynomial $P(z,\theta)$  can be
expanded over the above $Y_{b,N_s}$ as
\begin{equation}
P(z,\theta)=\sum_{N_s}Y_{b,N_s}A_{N_s}
\end{equation}
where the $A_{N_s}$ factors are given by [12]
(the index $s$ is omitted ):
$$A_k=\frac{k}{2}(u-v-\mu\nu)
\partial_z^2P(z,\theta)-\frac{k[P(v+\mu\nu,\nu)+
P(u+\nu\mu,\mu)]}{(u-v-\mu\nu)},$$
\begin{equation}
A_u=\frac{k}{1-k}\left(1-\frac{\mu\nu}{\sqrt k(u-v)}\right)
P(u,-\frac{1-\sqrt k}{2\sqrt k}\mu),
\end{equation}
$$A_\mu=\frac{\sqrt k}{2(1-\sqrt k)}
\left(1-\frac{\mu\nu}{\sqrt k(u-v)}\right)\partial_\theta
P(u+\theta\mu,\theta)-\frac{(\mu-\nu)A_u}{\sqrt k(u-v)}.$$
The $A_v$ and $A_\nu$ values can be obtained
from $A_u$ and $A_\mu$, respectively, by the
replacements $k\rightarrow 1/k,u\rightarrow v, v\rightarrow
u,\mu\rightarrow\nu$ and $\nu\rightarrow\mu$.

The $Y_{b,N_s}(t)$ and $Y_{a,N_s}(t)$ polynomials satisfy
the following relation:
\begin{equation}
Y_{b,N_s}(t)+Q_{\Gamma_{b,s}}^{2}(t)Y_{a,N_s}(t_s^b)
=Y_{a,N_s}(t)+Q_{\Gamma_{a,s}}^{2}(t)Y_{b,N_s}(t_s^a).
\end{equation}
To prove eq.(A5) we replace $t$ by $t_s^a$ in the
first of eqs.(39) and $t$ by $t_s^b$ in the second of
them.  For both $G_{gh}(t_s^a,t')$ and $G_{gh}(t_s^b,t')$
appearing on the right side of (39) we again use
(39). The resulting left sides in both equations in
(39) appear to be the same because
$\Gamma_{a,s}$ commutate with $\Gamma_{b,s}$. It leads to
(A5).

{}From (40) and using that
$D(t)g_s^p=Q_{\Gamma_{p,s}}^{-1}(t)\gamma_s^p,
D(t)\gamma_s^p=Q_{\Gamma_{p,s}}^{-1}(t)$ and that $\partial_z
=D(t)D(t)$, one obtain both $\partial_z
Y_{p,N_s}(t)$ and $D(t)Y_{p,N_s}(t)$ as
\begin{eqnarray}
\partial_z Y_{p,N_s}(t)=-[D(t)\ln Q_{\Gamma_{p,s}}(t)]
D(t)Y_{p,N_s}+2[\partial_z \ln Q_{\Gamma_{p,s}}(t)]Y_{p,N_s}(t)-
2\partial_{N_s}\ln Q_{\Gamma_{p,s}}(t)\nonumber \\
D(t)Y_{p,N_s}(t)=2Q_{\Gamma_{p,s}}^{-1}(t)
[D(t)Q_{\Gamma_{p,s}}(t)] Y_{p,N_s}(t)+
2Q_{\Gamma_{p,s}}(t)\partial_{N_s}\gamma_s^{(p)}(\theta,z).
\end{eqnarray}

\section{Green function  in the boson string theory}
In (44) the boson Green function  $R_b(z,z')$ is
given by [25]
\begin{equation}
R_b(z,z')=\sum_\Gamma\ln\left(\frac{[z-g_\Gamma(z')][-c_\Gamma
z^{(o)}+ a_\Gamma]}{[-c_\Gamma
z+a_\Gamma][z^{(o)}-g_\Gamma(z^{(1)})]}\right)
\end{equation}
$z^{(o)}$ and  $z^{(1)}$ being arbitrary constants. Furthermore,
[14] periods
$J_{(o)s}(z)$ of the above  $R_b(z,z')$ are given by
\begin{equation}
J_{(o)s}(z)={\sum_\Gamma}^\prime\ln\frac{z-g_\Gamma(u_s)}
{z-g_\Gamma(v_s)}\quad,
\end{equation}
$u_s$  $v_s$ being the fixed points of the  $\Gamma_s$
transformation.  In (B8) the summation
is performed over all the group products except those  with
the rightmost to be a power of the above $\Gamma_s$.
The period matrix $\omega^{(o)}$ turns out [14,26]
to be
\begin{equation}
\omega_{sp}^{(o)}={\sum_\Gamma}^{''}\ln\frac{[u_s-g_\Gamma(u_p)]
[v_s-g_\Gamma(v_p)]}{[u_s-g_\Gamma(v_p)][v_s-g_\Gamma(u_p)]}
+\delta_{ps}\ln k_s\quad.
\end{equation}
In (B9) the summation is performed over all  $\Gamma$
except those that have the leftmost to be a power of
$\Gamma_s$,  the rightmost being a power $\Gamma_r$.
Besides, $\Gamma\neq I$, if $s=p$.

\section{Integral conditions}
To derive both (50) and (52) we consider
the $F_{1/2,s}(t)$ function to be  1/2-supertensor
under both $\Gamma_{a,s}$ and $\Gamma_{b,s}$ mappings and
prove the following relation:
\begin{equation}
\int\limits_{C_s}F_{1/2,s}(t)dtK_s^{(1)}(t,t')=
-\int\limits_{C_b^{(s)}}F_{1/2,s}(t)dt
2\pi i\eta_s^{(1)}(t')+
\int\limits_{C_b^{(s)}}F_{1/2,s}(t)dt\varphi_s(t)f_s(t')
\end{equation}
where $dt=d\theta dz/ 2\pi i$ and the $C_s$ contour  surrounds
both $C_s^{(-)}$ and $C_s^{(+)}$ circles (26)
together with the $\tilde C_s$ cut  arising for $l_{1s}\neq0$.
To prove (C10) it is convenient to use the $t_s$
supercoordinates (23) instead of $t$ to be variables
of the integration. Moreover, we choose the intersect
$z_s^{(+)}=\tilde C_s\bigcap \hat
C_s^{(+)}$ as $z_s^{(+)}=g_s^{(b)}(z_s^{(-)})$ where $z_s^{(-)}=
\tilde C_s\bigcap \hat C_s^{(-)}$ and
$z_s\rightarrow g_s^{(b)}$ is
the Schottky transformation corresponding to $2\pi$-twist about
$B_s$-cycle.  The integral along the $\tilde C_s$ cut disappears
owing to (29).
The integral along  $\hat C_s^{(+)}$ is reduced to
the integral along
the $\hat C_s^{(-)}$ contour by the Schottky transformation
(17).  After the above integral to be summed with
the integral along $\hat C_s^{(-)}$, the right side of
(C10) arises.

Taking $F_{1/2,s}=f_s$, one obtains that the left side of
(C10) is equal to $f_s$ owing to the first of
eqs.(51). So two equations in
(50) appear. To prove the third relation in
(50) we choose $F_{1/2,s}(t')$ to equal to
$K_s^{(1)}(t,t')$. In this
case the integral along $C_s$-contour  also can be reduced to
the integral along $C_b^{(s)}$, but  the integrand appears to
be different from that given in (C10) because
$K_s^{(1)}(t,t')$ has periods about $B_s$-cycle. Owing to the
second equation of eqs.(51), one obtains a
number of relations, both the third equation of eqs.(50)
and (52) being among of them.

As far as the $\Gamma_{a,s}$ and $\Gamma_{b,s}$ mappings
are the same for both $K_s^{(1)}$ and $K$, the above
consideration can be applied also to the integration over
$t_1$ along the $C_s$ contour of $f_s(t)K(t_1,t')$.
In this case the first term on
the right side of (C10) contains $\eta_s$ instead of
$\eta_s^{(1)}$.  The last term disappears because the fermion
zero mode is absent in the case discussed. So the last of
eqs.(50) remains to be true, if one takes  $K$ instead
of $K_s^{(1)}$. In addition, one concludes that
\begin{equation}
\int\limits_{C_s}f_s d\theta dz K_(t,t')=0
\quad{\rm and}\quad
\int\limits_{C_s}K(t',t)d\theta dz\varphi_s(t)=0.
\end{equation}
The second eq.(C11) follows from the first one
because $f_s(t)=D(t)\varphi_s(t)$ and
$D(t)R_s^{(1)}(t,t')=K_s^{(1)}(t',t)$.

To prove  eq.(53) we write down the sum over $r$ in
(53) as
\begin{equation}
-\sum_{r=1}^{n}\int\limits_{C_r}K_{(o)}(t,t_1)
dt_1 K_r^{(1)}(t_1,t_2)dt_2K(t_2,t')+
\int\limits_{C_r}K_{(o)}(t,t_1)
dt_1 K_{(o)r}^{(1)}(t_1,t_2)dt_2K(t_2,t')
\end{equation}
where one can think  that in the integral over $z_1$ the pole at
$z_1=z_2$ in both $K_{(o)r}^{(1)}(t_1,t_2)$ and
$K_r^{(1)}(t_1,t_2)$ is surrounded by $C_r$-contour. When the
integral over $z_1$ in the second term
is reduced to $C_b^{(r)}$, the contribution of the above pole
appears  to be
\begin{equation}
\sum_{r=1}^n\int\limits_{C_r}K_{(o)}(t,t_2)
dt_2K(t_2,t')= K(t,t')-K_{(o)}(t,t').
\end{equation}
The rest of (C12) together with to two other terms on
the right side of
(53) turns out to be equal to zero owing to (C11).
It proves  eq.(53).

To prove (54) and (55) we do as in proving
(C10).  Only the integral along the cut
contributes into the right side of  (54).
The eq.(55) can be obtained also by the calculation of
the period about $B_s$-cycle of $J_s$ given by (54).

The kindred statements given in Sec.V can be derived in the
quite similar manner. For this aim we consider the
$F_{3/2,s}(t)$ function to be  3/2-supertensor under both
$\Gamma_{a,s}$ and $\Gamma_{b,s}$ mappings and prove the
following relation:
\begin{equation}
\int\limits_{C_s}F_{3/2,s}(t)dt \tilde
G(t,t')=- \sum_{\{N_s\}}\left[\int\limits_{C_b^{(s)}}
F_{3/2,s}(t) dt
F_{b,N_s}(t) +\int\limits_{C_a^{(s)}}F_{3/2,s}(t)dt
F_{a,N_s}(t)\right] \Delta_{N_s}(t')
\end{equation}
where the $C_b^{(s)}$ contour and the $C_a^{(s)}$ path are the
same as in (66).
mappings.  Moreover, $\tilde G_s(t,t')$ is the Green function
that transforms under the $(\Gamma_{a,s},\Gamma_{b,s})$
mappings by (63) with $F_{b,N_s}(t)$
superfunctions instead of the polynomials
and $\Delta_{N_s}(t')$ to be 3/2-supertensors instead of
$\chi_{N_s}(t')$ in (63).

Being applied for $F_{3/2,s}(t)=G(t_1,t)$ together with
$\tilde G(t,t')$ to be $G(t,t')$, one obtains that the left side
of (C14) is equal to zero that leads to  eq.(66).
To prove (73) one  uses (C14) for
$F_{3/2,s}(t)=\theta G_f(z_1,z)$ and $\tilde G(t,t')=
\theta G_{(\sigma)}(z,z')$.
To prove (76) one  uses (C14) for
$F_{3/2,s}(t)=S_\sigma (t_1,t)$ and $\tilde G(t,t')=
G(t,t')$.
To prove (79) one  uses (C14) for
$F_{3/2,s}(t)=\theta G_{(\sigma)}(z_1,z)$ and $\tilde G(t,t')=
\theta S_{(f)\sigma}(z,z')$.
Moreover, employing
$F_{3/2,s}(t)=S_{\sigma,s}^{(1)}(t_1,t)$ and $\tilde G(t,t')=
S_{\sigma,s}^{(1)}(t,t')$, one obtains that
\begin{equation}
\int\limits_{C_b^{(s)}}S_\sigma(t,t') \frac{d\theta'
dz'} {2\pi i}\hat Y_{\sigma,N_s}^{(1)}(t')=0.
\end{equation}
In Sec.VI the above method is used to derive the integral
relations (87) for $\tilde\chi_{N_s}(t)$ and
eqs.(88) for $\chi_{N_s}(t)$ and
$\Psi_{\sigma,N_s}(t)$.

\section{ Integral representations }
To obtain eq.(91)
we represent $\partial_{N_r} J_m(t)$ as
it follows
\begin{equation}
\partial_{N_r} J_m(t)=-\int\limits_{C(z)}K(t,t')dt'
\partial_{N_r} J_m(t')=\sum_r\int\limits_{C_r}K(t,t')dt'
\partial_{N_r} J_m(t')
\end{equation}
where $K(t,t')$ is defined by eq.(36) and $dt=d\theta
dz/2\pi i$.  The integrals along $C_r$ are
reduced to the integrals  along both the $C_a^{(r)}$
and $C_b^{(r)}$ paths defined  by (66).  For the
calculation of the periods of $\partial_{N_r} J_m(t)$
we use (92).
We use also  the second of eqs.(50)  for
$K(t,t')$.   The resulting
relations are found to be
\begin{equation}
\partial_{N_r} J_m(t)=-\pi
i\sum_{p=a,b}\int\limits_{C_p^{(r)}}K(t,t')
dt\{[D(t')\eta_m(t')]Y_{N_r}(t')-
D(t')[\eta_m(t')Y_{N_r}(t')]\}
\end{equation}
where $\eta_m(t')$ are the half-forms (37).
Calculating the periods about $B_s$-cycle of the left
and right sides of (D17), we obtain the  integral
representation for $\partial_{N_r}\omega_{mn}$
in the following form:
\begin{equation}
\partial_{N_r}\omega_{mn}=-\pi
i\sum_{p=a,b}\int\limits_{C_p^{(r)}}\eta_n(t)
dt\{[D(t')\eta_m(t')]Y_{N_r}(t')-
D(t')[\eta_m(t')Y_{N_r}(t')]\}.
\end{equation}
The last term in (D18) is integrated by parts.  Out
integral terms disappear owing to (A5). And we obtain
eq.(91).

To obtain another  useful relations, one can integrate
along the  $C_r$ contour the quantities
$f_r(t')\partial_{N_r}K_r^{(1)}(t',t)$ and $f_r(t)
\partial_{N_r}\varphi_r(t)$.
The above integrals are equal to zero
because  both the integrands have no singularities
outside of the $C_r$ contour.
In this case, using (92), we
obtain that
\begin{eqnarray}
\sum_{p=a,b}\int\limits_{C_p^r}f_r(t')dt'E_{N_r}^{(0)}
\left(K_r^{(1)}(t',t)\right)-
\int\limits_{C_b^r}f_r(t')dt'\left[E_{N_r}^{(0)}
\left(\varphi_r(t')f_r(t)\right)-
\partial_{N_r}[\varphi_r(t')f_r(t)]\right]=0\nonumber \\
{\rm and}\quad\sum_{p=a,b}\int\limits_{C_p^r}f_r(t')dt'
E_{N_r}^{(0)}\left(\varphi_r(t')\right) =0
\end{eqnarray}
where $E_{N_r}^{(0)}\left(F(t,t')\right)$
is defined by (93) at $q=0$.

The value of $\partial_{N_r}K_r^{(1)}(t,t')$
for the even genus-1
spin structures is given in Sec.VI.
For the odd genus-1 spin structure the same method gives that
\begin{eqnarray}
\partial_{N_r}K_r^{(1)}(t,t')=\int\limits_{C_a^r}
K_r^{(1)}(t,t_1)dt_1E_{N_r}^{(0)}\left(K_r^{(1)}(t_1,t')\right)
-\varphi_r(t)\int\limits_{C_b^r}f_r(t_1)dt_1
\partial_{N_r}K_r^{(1)}(t_1,t')+\nonumber \\
\int\limits_{C_b^r}\left[K_r^{(1)}(t,t_1)+\varphi_r(t)f_r(t_1)
\right]dt_1
\left[\partial_{N_r} \left(\varphi_r(t_1)f_r(t')\right)+
E_{N_r}^{(0)}\left(K_r^{(1)}(t_1,t')-
\varphi_r(t_1)f_r(t')\right)\right]
\end{eqnarray}
The same method allows also to obtain the following
representation for
$\partial_{N_r}\varphi_r(t)$:
\begin{eqnarray}
\partial_{N_r}\varphi_r(t)=\sum_{p=a,b}\int\limits_{C_p^r}
K_r^{(1)}(t,t_1)dt_1E_{N_r}^{(0)}
\left(\varphi_r^{(1)}(t_1)\right)+
\varphi_r(t)\int\limits_{C_b^r}f_r(t_1)
E_{N_r}^{(0)}\left(\varphi_r^{(1)}(t_1)\right)-\nonumber \\
\varphi_r(t)\int\limits_{C_b^r}f_r(t_1)
\partial_{N_r}\varphi_r(t_1).
\end{eqnarray}

To obtain $\partial_{M_r}S_{\sigma,r}^{(1)}(t,t')$ we integrate
along the infinitesimal $C(z_1)$ contour the quantities
$S_{\sigma,r}^{(1)}(t,t_1)
\partial_{M_r}S_{\sigma,r}^{(1)}(t_1,t')$
and we deform the contour as it was
explained above. The result is
\begin{eqnarray}
\partial_{M_r}S_{\sigma,r}^{(1)}(t,t')=
\sum_{p=a,b}\int
\limits_{C_p^r}S_{\sigma,r}^{(1)}(t,t_1)dt_1
E_{M_r}^{(-2)}\left(S_{\sigma,r}^{(1)}(t_1,t')\right)+
\sum_{N_r}\int\limits_{C_b^r}
S_{\sigma,r}^{(1)}(t,t_1)dt_1\times
\nonumber \\ E_{M_r}^{(-2)}
\left(S_{\sigma,r}^{(1)}(t_1,t')\right)-
\sum_{N_r}\int\limits_{C_b^r}S_{\sigma,r}^{(1)}(t,t_1)dt_1
\partial_{M_r}[\hat
Y_{\sigma,N_r}^{(1)}(t_1)\Psi_{\sigma,N_r}^{(1)}(t')]
\end{eqnarray}
where $E_{M_r}^{(-2)}\left(F(t,t')\right)$ is
defined by (93) at $q=-2$. Furthermore, one could
integrate over $t$
along the above $C(z_1)$ contour the quantities
$\chi_{N_s}(t)\partial_{M_r}S_\sigma(t,t')$. In this case one
obtains that
\begin{eqnarray}
\sum_{p=a,b}\int\limits_{C_p^r}\chi_{N_s}(t)dt
E_{M_r}^{(-2)}(S_\sigma(t,t'))+
\sum_{N_r}\int\limits_{C_b^r}\chi_{N_s}(t)dt\times
\nonumber \\ E_{M_r}^{(-2)}\left(S_\sigma(t,t')\right)-
\sum_{N_r}\int\limits_{C_b^r}\chi_{N_s}(t)dt
\partial_{M_r}[\hat
Y_{\sigma,N_r}^{(1)}(t_1)\Psi_{\sigma,N_r}(t')]=0.
\end{eqnarray}

\section{Derivatives of the Schottky group multipliers.}
To obtain the desired values of $\partial_{u_s}k,
\partial_{v_s}k$ and of $\partial_{k_s}k$, we write down
the Schottky group product $g$ having the  multiplier to be
$k$ as follows
\begin{equation}
g=\prod_{j=1}^q g_{(j)}g_s^{n_j}
\end{equation}
where $g_s^{n_j}$ are powers of the given $g_s$ basic
Schottky group element and the $g_{(j)}$ group products do
not contain $g_s$. In this case the $\delta g(z)$ alteration of
$g(z)$ with respect to  variations $\delta u_s$, $\delta v_s$
and $\delta k_s$ of $k_s,u_s$ and  $v_s$
is given by
\begin{equation}
Q_g(z)^2\delta g(z)=\sum_{j=1}^q
Q_{g(>j)}^2(z)[Y_{k_s}(z_j)\delta k_s+\frac{(1-k_s^{n_j})k_s}
{k_s^{n_j}((1-k_s)}Y_{u_s}(z_j)\delta u_s+
\frac{(1-k_s^{n_j})}
{((1-k_s)}Y_{v_s}(z_j)\delta v_s]
\end{equation}
where $Q_g^{-2}(z)=\partial_z g(z)$, $z_j=g(>j;z)$ and
the $g(>j)$ mapping is defined as
\begin{equation}
g(>j)=\prod_{p=j+1}^q g_{(p)}g_s^{n_p}
\end{equation}
with $g(>q)=I$. Furthermore, $Y_{u_s},Y_{v_s}$ and $Y_{k_s}$
in (E25) are none other than the $Y_{b,N_s}$
polynomials defined by (40) at zero odd parameters.  On the
other hand, from (40) it is follows also that
\begin{equation}
Q_g(z)^2\delta g(z)=Y_k(z)\delta k+ Y_u(z)\delta u+
Y_v(z)\delta v
\end{equation}
where $Y_u(z), Y_v(z)$ and $ Y_k(z)$ are the degree-2
polynomials corresponding to the $z\rightarrow g(z)$
transformation. Moreover,  the right side of
(E25) can be expanded over the above $Y_u(z), Y_v(z)$ and $
Y_k(z)$ polynomials by means of eqs.(A3). In this case one
obtains both $\delta u,\delta v$ and $\delta k$ in the terms of
$\delta u_s, \delta v_s$ and $\delta k_s$. So, one can  obtain the
derivatives with respect to $ u_s,  v_s$ and $ k_s$ of every $u,v$
and of $k$.

\end{document}